# Boosting high-current alkaline water electrolysis and carbon dioxide reduction with novel CuNiFe-based anodes

Nusrat Rashid[a], Shurui Yang[a], Galyam Sanfo[a], Isabelle Ewing[a], Zahra Ibrahim Albu[a], Xinjuan Li[b], Tianhao Wu[b], Prajna Bhatt[c,d], Mathieu Prevot[e], Laurent Piccolo[e], Mahmoud Zendehdel[f], Robert G. Palgrave[c], Caterina Ducati[b], Mojtaba Abdi-Jalebi[a*]

[a] *Institute for Materials Discovery, University College London, London, WC1E7JE, UK*

[b] *Department of materials Science and Metallurgy, University of Cambridge, Cambridge, CB30FS, UK.*

[c] *Department of Chemistry, University College London, London, WC1H0AJ, UK.*

[d] *Istituto Officina dei Materiali (IOM)-CNR, Laboratorio TASC, in Area Science Park, S.S.14, Km 163.5, Trieste I-34149, Italy*

[e] *Université Claude Bernard Lyon 1, CNRS, IRCELYON, Villeurbanne, F-69100, France.*

[f] *Iritaly Trading Company S.R.l., Via Volturno 58, 00185 Rome, Italy*

[*]*Email:* m.jalebi@ucl.ac.uk

**Abstract:** The transition to a green hydrogen economy demands robust, scalable, and sustainable anodes for alkaline water electrolysis operating at industrial current densities (>1 A/cm²). However, achieving high activity and long-term stability under such conditions remains a formidable challenge with conventional catalysts. Here, we report a novel trimetallic CuNiFe anode fabricated through a rapid, single-step electrodeposition process at room temperature without organic additives. The catalyst exhibits an exceptionally low overpotential of <270 mV at 100 mA cm$^{-2}$ and operates stably for over 500 hours at 1 A cm$^{-2}$ in 30 wt% KOH. In a practical anion exchange membrane water electrolyzer (AEM-WE), the CuNiFe anode enables a current density of 2.5 A cm$^{-2}$ at only 2.5 V, with a voltage efficiency of 66.8%. Beyond water splitting, this anode also significantly enhances $CO_2$ electrolysis, tripling the $CO_2$ reduction current density and steering selectivity toward valuable multi-carbon products when paired with commercial copper cathodes. A cradle-to-gate life cycle assessment confirms that the CuNiFe anode reduces the carbon footprint by an order of magnitude and decreases environmental impacts by 40–60% across multiple categories compared to benchmark IrRuO$_2$. Our work establishes a scalable, high-performance, and environmentally benign anode technology, paving the way for cost-effective electrochemical production of green hydrogen and carbon-neutral chemicals.

**Introduction**

Electrochemical oxygen evolution plays a critical role in alternate-energy schemes as it provides the electrons and protons needed for converting electricity into reduced chemical fuels like hydrogen or hydrocarbons through artificial photosynthesis.[1,2] Indeed, the oxygen evolution reaction (OER) is the simplest noninterfering half-cell reaction that can be paired with a reduction process, such as the hydrogen evolution reaction (HER) or the $CO_2$ reduction reaction in electrolysers.[3] However, the OER is energy intensive as it requires a significant overpotential (in addition to an already high reversible oxidation potential) to reach substantial current densities, occurs in conditions prone to trigger anodic corrosion (in particular at industrial scales, where current densities >100 mA/cm$^2$ are required), thereby increasing the manufacturing cost as well as operating cost of electrolysers.[4–6] The complex formulation of catalysts, scaling of process and materials add further complications and cost to a practical solution for alkaline electrolysers.[7,8] Therefore, to enable industrial deployment of alkaline electrolysis, it is crucial to develop inexpensive, environmentally benign electrocatalysts with a low carbon footprint and high resilience at multi-ampere scales.[9]

The urgency of achieving net-zero has driven extensive research in OER optimization, leading to significant advancements in catalyst development, such as noble metal oxides,[9] layered double hydroxides (LDHs), oxyhydroxides,[10,11] phosphides, molecular catalysts[12], first row transition metal based (Ni, Co, Mn, Fe, Cu, etc and their bimetallic, trimetallic compositions),[13–16] and sulphides.[17] Among them NiFe based LDHs are usually considered as prototypal state-of-the-art non-noble electrocatalyst at the lab-scale, as they show excellent activity in alkaline conditions on account of their abundant active sites and unique lamellar structure yielding overpotentials of 200-250 mV @10mA/cm$^2$ current density for the OER.[18–21] Still, despite these advancements only few electrocatalysts can replace existing metal-electrodes like Raney nickel in industry set-up due to low stability at ampere currents, higher corrosion leading to failure, and complicated upscaling fabrication processes unfit for existing industrial infrastructure. The main concerns include mass transfer limitations under high current densities,[22,23] anodic corrosion of anode during violent gas release and simultaneous heating,[24,25] dissolution of binder accompanying catalyst breakdown,[26–28] and cost-effective deposition of catalysts at larger scale.[29] Therefore, a substantial technological gap remains to be bridged for Ni-based LDH electrocatalysts to be included in industrial electrolysers.

Herein, we report a single-step room conditions electrodeposition of copper rich trimetallic anode (CuNiFe) from slightly acidic metal solution, which achieved highly active and stable oxygen evolution for 500 h at 100 mA/cm$^2$, and 20 h at 1 A/cm$^2$ in 1M KOH solution in an H-cell. In 30% KOH the catalyst was stable for 500 h at 1A/cm$^2$ in AEM-WE with cell potential of 1.85 V and no activity attenuation for 6 hours at 2.5A/cm$^2$. The catalyst prepared in 5 minutes has a unique conformal, binder-less deposition of Cu core, NiFeCu top layer, accelerating electron transfer whilst resisting the alkaline dissolution amplified by high operational potential. The catalyst is easily scalable with retention of all properties and shows zero degradation under 30% KOH and 60-80 ºC temperatures for 24 hours.

**Results**

**Synthesis and structure of CuNiFe**

In a 3-electrode cell, a cleaned 2 cm * 2 cm nickel foam (Ni-Foam) cathode was used to electrodeposit CuNiFe from a metal precursor solution consisting of Cu$^{2+}$, Fe$^{3+}$, and Ni$^{2+}$ (5 mM each), along with 30 mM of H$_2$SO$_4$. A platinum (Pt) anode and an Ag/AgCl reference electrode were employed, and the deposition was carried out via ON current pulses of -100 mA/cm$^2$ amplitude for 0.5 sec and OFF pulses of 0 mA/cm$^2$ amplitude for 0.05 sec for a total deposition time of 150 seconds (supplementary fig. 1). The colour of the nickel foam turns reddish brown indicating deposition of CuNiFe (**Fig. 1a**). The CuNiFe was optimised for its nanostructure and electrical properties by tailoring different synthesis parameters (supplementary note 1, 2, and 3 and supplementary fig 2-8) and analysing the electrochemical response. The crystalline phases of the optimised CuNiFe catalyst and Ni-foam analysed by X-ray diffraction (XRD) are shown in Supplementary Fig. 9. The XRD of CuNiFe on Ni foam were reference corrected to gain insights into the crystalline structure of the film. **Fig. 1b** shows the peaks for NiCuFeO$_2$ phase at 13º, and 35º in addition to peaks for Cu (0), Ni-Fe LDH.[30]

Systematic structural characterisation like scanning electron microscopy (SEM) shows a homogenous repetitive grain structures deposited on the Ni foam (**Fig. 1c**). Uniform elemental distribution of Cu, Ni, Fe and O was assessed by Energy-dispersive X-ray (EDS) spectroscopy (Supplementary Fig. 10). At higher magnifications nanoparticles can be seen covering conformally the surface of nickel foam without any pinholes (Supplementary Figs. 11). Moreover, the particles are adhered compactly to substrate and other neighbouring particles without formation of micron size clusters.

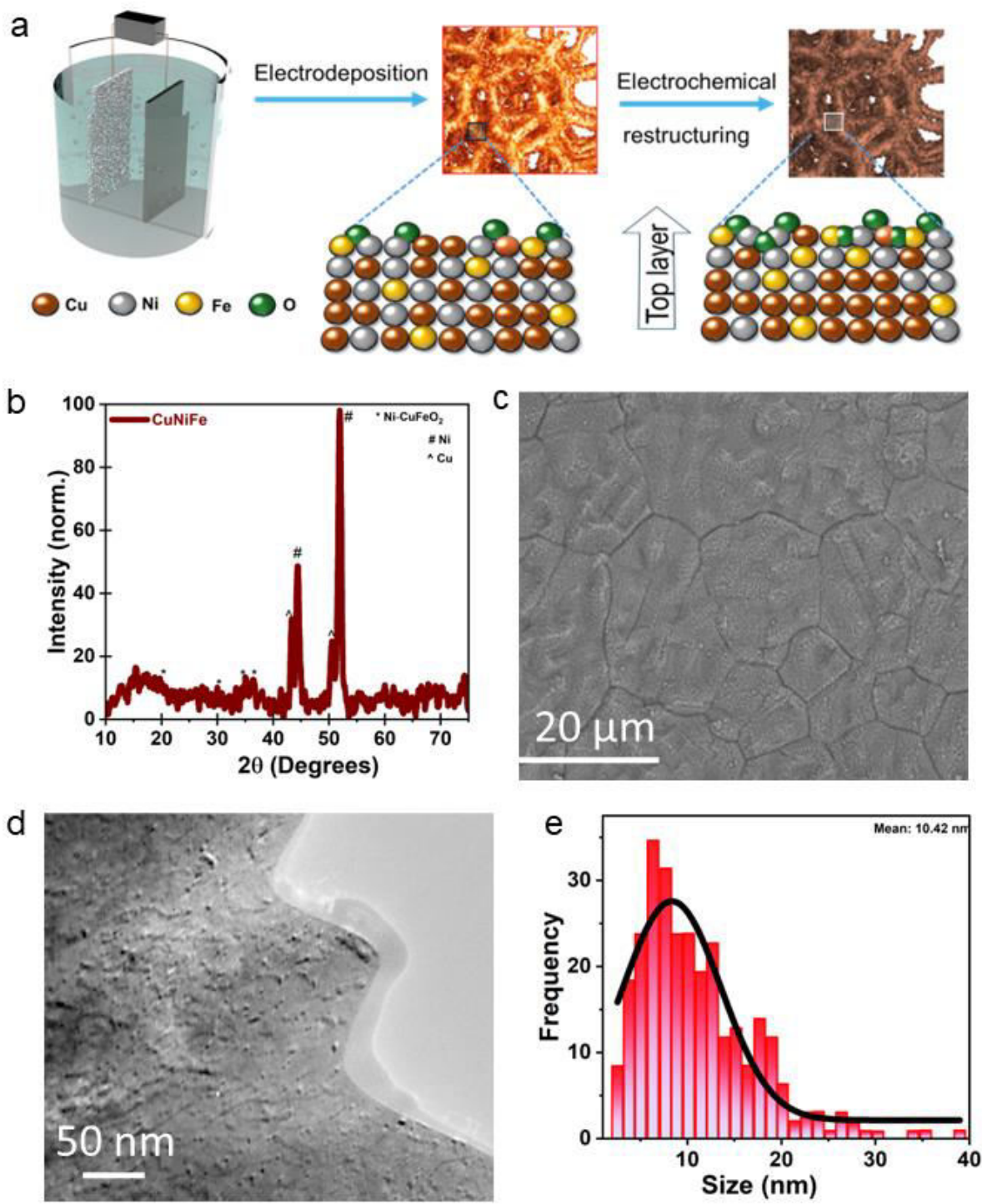

***Fig. 1: Synthesis and structural characterisation of CuNiFe. a**, Schematic for one-step electrodeposition of the catalyst film and its electrochemical activation through surface reorganisation. **b**, XRD graph showing peaks corresponding to the phases Ni-CuFeO$_2$, Cu and Ni. **c**, SEM image showing conformal, and periodic grains of the catalyst film. **d**, TEM image of the catalyst demonstrating homogenous composition of spherical nanoparticle. **e**, Size distribution and mean size of the particles calculated from TEM.*

Furthermore, high resolution transmission electron microscopy (HRTEM) images of the nanostructured grain extracted from freshly prepared CuNiFe sample via focussed ion beam (FIB) shows the distribution of the particles and their sizes (**Fig. 1d**). The particle size ranges from 5-20 nm with a frequency of 20, and a mean particle size of ~11nm (**Fig. 1e**). The Raman spectra (Supplementary Fig. 12) show peaks for CuFe at 129, 150 cm$^{-1}$, Cu-O at 295 cm$^{-1}$, and NiFe-Cu at 547, and 628 cm$^{-1}$.[31–33] A Raman peak attributed to internal *Eg* mode in Fe (III) is also present at 413 cm$^{-1}$. [34,35]

The structural and chemical stability is one of the key parameters for sustainable catalytic activity.[36,37] High resolution TEM images (**Fig. 2 a**) show spherical nanostructure of the deposited film. Multiple sites were analysed to assess the phase uniformity on the surface by applying fast Fourier transformation on HRTEM images. The phases analysed through diffraction pattern were $NiCuFeO_2$ and CuNiFe-LDH as shown in **Fig. 2 b**. The high angle annular dark filed (HAADF) image (Supplementary Fig. 13) was used to determine the elemental distribution in freshly prepared CuNiFe. **Fig. 2c-e** shows the distribution of copper mostly in core with Fe and Ni rich outer layer. As demonstrated, most of the Ni is in top layers attributed to lagged deposition of Ni. The energy dispersive spectroscopic (EDS) graphs for Cu, Ni, and Fe are shown in Supplementary Fig. 14. Ni being most prevalent in the observed layers (some contribution from the substrate), followed by Cu and Fe as expected. The presence of Fe in core can be attributed to the competitive deposition of Cu and Fe. However, Ni rich outer surface can be attributed to optimised pulses that allow Ni deposition on depopulation of copper ions and fading drag power of ON pulse.

From the XPS analysis Ni exists as Ni (II) with a binding energy value of 855.5eV,[38,39] and Fe 2*p* is complex to discern due to the overlapping region at the Al Kα excitation energy between the core level and Ni LMM Auger peaks[39,40], is most probably present as a mixture of (II) and (III) oxidation states (**Fig. 2 d-e**). Copper (**Fig. 2c**) is in its native state (binding energy 932.6eV) forming a metal core high in conductance and a seat of on-demand Lewis basicity.[41] The atomic ratio of Cu, Ni, Fe in Ni-foam and CuNiFe from XPS analysis were determined as 0, 94.4, 5.6% and 79.3, 16, 4.7%, respectively (**Table S1**). The presence of Fe in lower concentrations have proven to enhance the activity, especially in higher oxidation states without high peroxidising dissolution.[42–44] Smaller, impurity level (~5%) Fe is also confirmed in CuNiFe by TEM-EDS. The XPS survey scan (Supplementary Fig. 15 a) confirms presence of Cu, Ni, Fe along with some O and C in the CuNiFe film. From C1s and O1s (Supplementary Fig. 15 b-c) shows M-OH and M-O bonds in both Ni-foam and CuNiFe.

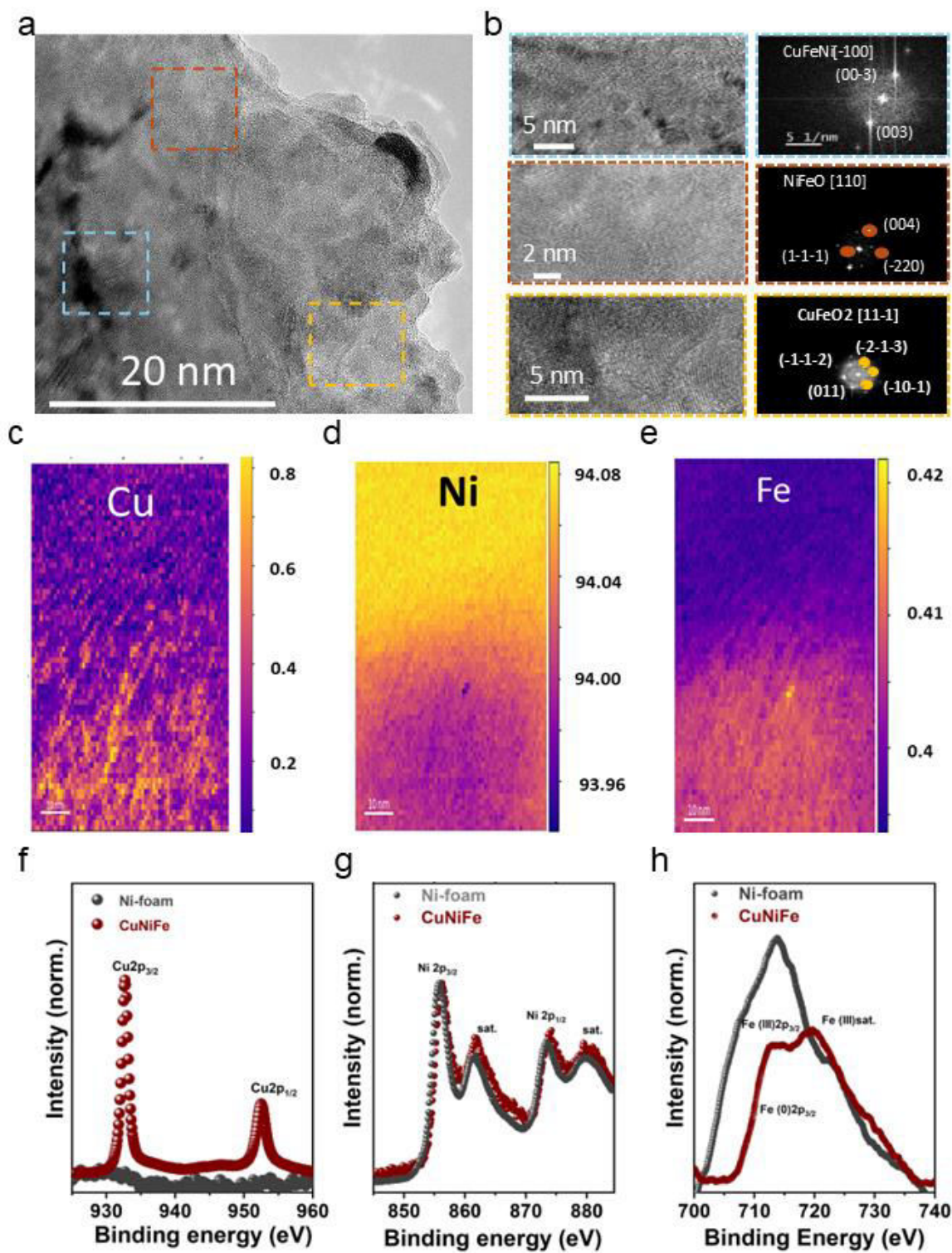

***Fig. 2: Microscopic and electronic structure of CuNiFe. a,*** *HR-TEM of the catalyst film,* ***b,*** *showing presence of CuFeO$_2$ and Ni-CuFeO$_2$ and NiFeO in diffraction patterns created using FFT from HR-TEM. Elemental distribution and mapping of Cu **(c)**, Ni **(d)**, and Fe **(e)** in the freshly prepared CuNiFe. Baseline corrected XPS core level spectra of,* ***(f)*** *Cu 2p,* ***(h)*** *Ni 2p, and **(h)**Fe 2p (overlapping with Ni auger peaks) in CuNiFe and Ni-foam.*

**Electrocatalytic performance toward oxygen evolution**

The OER performance of as-synthesised electrocatalysts was assessed in a 3-electrode H-cell set-up in 1M KOH solution at room conditions. The 80% iR corrected cyclic voltammetry (CV) polarization curve of CuNiFe shows faster oxygen evolution compared to Ni-foam, $IrO_2$.GDE, and Ni mesh [45] (**Fig. 3a**). An overpotential of 270 mV at 100 mA/cm$^2$ current was observed on CuNiFe, which is 130 mV lower than Ni mesh. Ni foam and $IrO_2$ need more than 1.95 V to reach current density of 100 mA/cm$^2$. The geometric activity trackers are often over-estimated[46] as electrochemical active sites are usually higher, and the current can increase with decrease in overpotential or increased mass loading of catalyst.[47] Therefore, it is recommended to use current normalised to electrochemically active surface area (ECSA), mass activity, and substrate baseline correction to depict the intrinsic activity.[46,48] To assess the ECSA we recorded cyclic voltammetrograms at various (50-300 mV/s) scan rates to determine the diffusion layer capacitance (Supplementary fig 3). Supplementary Fig. 16 shows the ECSA, mass loading, and substrate current normalised CV polarization curves for CuNiFe. A high mass activity of 80 A/g was achieved at 1.8 V, calculated using the average catalyst mass deposited on the nickel foam substrate (Supplementary table S2).

The validity of the substrate current correction and the intrinsic activity of CuNiFe were further confirmed by depositing the material on an inert titanium (Ti) metal foil, which exhibits no oxygen evolution reaction (OER) activity itself.[49,50] The similar activity observed at lower potentials for CuNiFe on both Ni-foam and Ti foil (Supplementary Fig. 2 and note 1) demonstrated the negligible contribution of the substrate to the measured current. Consequently, the current from the bare nickel foam was subtracted from the CuNiFe data to isolate the total current originating solely from the catalyst layer (supplementary fig. 16).

Electrochemical Impedance Spectroscopy (EIS) analysis of CuNiFe shows reduced charge transfer resistance on the interface compared to Ni-foam (**Fig. 3b**). Copper concentrating in the core forms a high conduction interlayer which increases the overall charge transfer and decreases contact resistance, and charge transfer resistance. Overpotentials of 270, 280, and 300 mV at 100, 200, and 500 mAcm$^{-2}$ respectively were achieved without any substantial activity attenuation. All overpotentials are referenced to the actual steady state chronopotentiometric polarization traces unlike most reported CV overpotentials which is transient in nature and does not yield similar results as observed in chrono methods.[51]

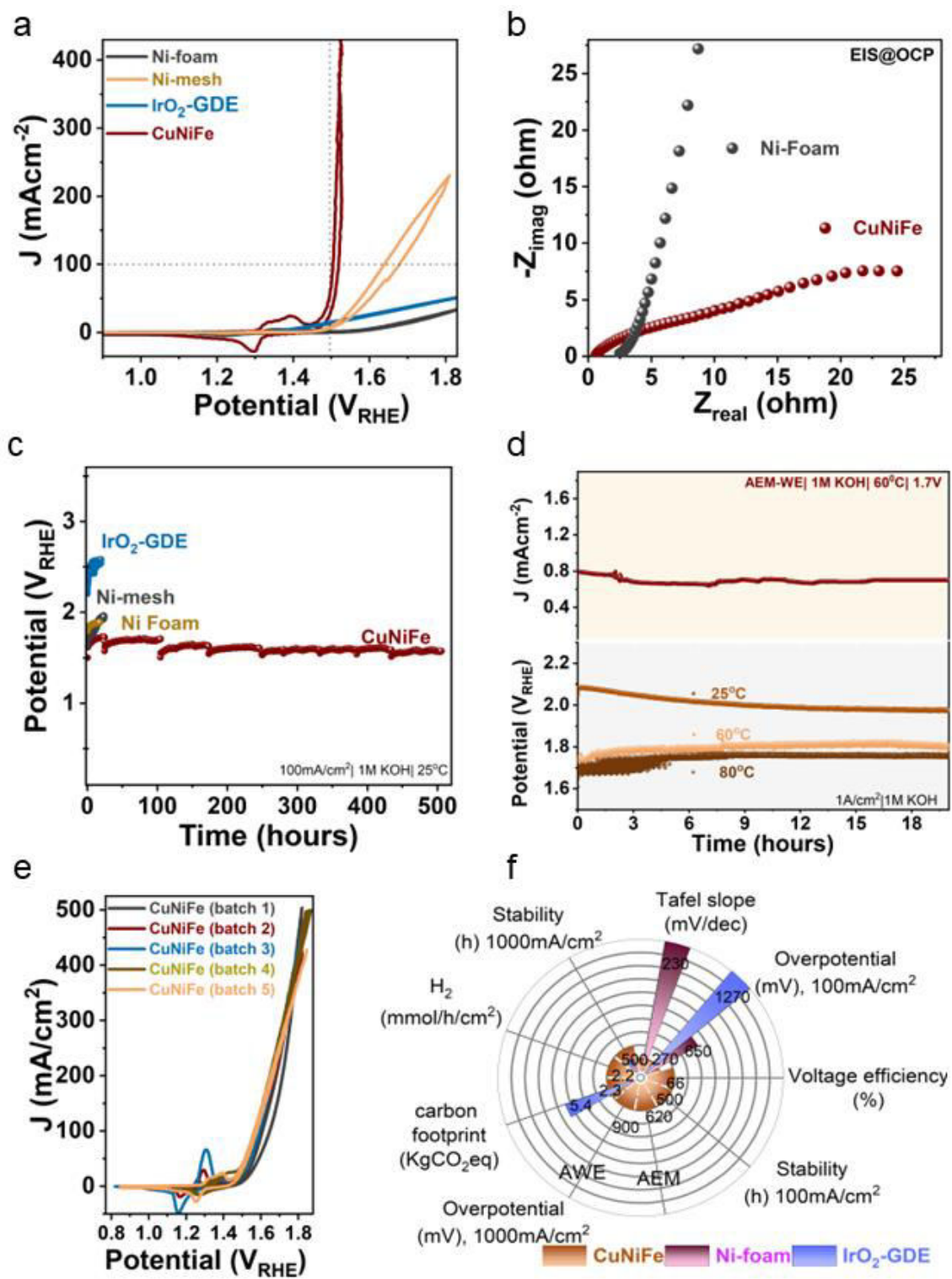

*Fig. 3: Electrochemical OER on the CuNiFe films in 1M KOH. **a**, 80% iR corrected CV traces of CuNiFe compared to Ni foam, Ni-Mesh, IrO$_2$-GDE. **b**, EIS of CuNiFe and Ni foam at OCP. **c**, stability of CuNiFe in 1M KOH at 100mA/cm$^2$ for 500 hours. **d**, chrono potentiometric traces of CuNiFe at 1 A/cm$^2$ current and different temperatures (25, 60, 80 °C) in H-cell. **e**, Repeatability and reproducibility of the catalyst film, 5 independently deposited films showing similar CV behaviour and currents. **f**, Activity plot of the CuNiFe at different currents, temperatures, and testing conditions.*

The durability of CuNiFe was tested at 100 mA/cm$^2$ for 500 hours in 3-electrodes H-cell employing a Sustainion® anion exchange membrane. An overpotential of ~280 mV was recorded at 100 mA/cm$^2$ as shown in **Fig. 3c** with <10 μV/hr voltage degradation**.** The H$_2$ and O$_2$ gases were quantified with an online gas chromatography. The faradaic efficiency was recorded as 87 and 102% (Supplementary Fig. 17 a-b) with an average gas production at 100 mA/cm$^2$ as 8.8 and 20.4 mmol/h.cm$^2$ for O$_2$ and H$_2$, respectively. In a H-cell and 1M KOH, CuNiFe showed cell potential of 2.08, 1.88, and 1.8 V at 1000 mA/cm$^2$ at 25, 60, and 80°C (**Fig. 3d)** with no substantial degradation for 20 hours of continuous operation. In a zero-gap configuration with nickel foam as cathode, and CuNiFe as anode and Sustainion® anion exchange membrane alkaline exchange membrane water electrolysis (AEM-WE) was performed at 60°C in 1M KOH. **Fig. 3d** shows a current of 380 mA/cm$^2$ at a cell potential of 1.7 V and stable operation for 20 hours. The ICP-MS analysis was done on collected electrolytes at intervals of 100 hours of continuous OER on CuNiFe for 500 hours. Supplementary Fig. 18 shows the net dissolution of Cu, Ni, and Fe as 35, 12, and 16 ppm respectively in 500 hours. We synthesised more than 10 CuNiFe in different batches and by different contributing researchers to test the repeatability of the process in same and different lab. **Fig. 3e** shows the CV recorded on CuNiFe prepared in five different batches in same and collaborator's lab. The catalyst shows promising repeatability of process and activity, much demanded by the commercialisation. There is less than +/- 30 mA/cm$^2$ current difference with each batch suggesting reliable repeatability of the film. **Fig. 3f** shows the activity, stability, Tafel slope, voltage efficiency at different currents, temperatures, and electrolyser type and configurations on CuNiFe.

The kinetic parameters of the CuNiFe anode were obtained from corresponding Tafel analysis from plotting overpotentials against log (J) (Supplementary Fig. 19). The as prepared CuNiFe catalysts show a 86 mV/dec Tafel slope compared to 230 mV/dec shown on Ni-foam. A strong adherence to the substrate and stable chemical phases at interface in prepared CuNiFe resisted any substantial microstructural degradation. The post-electrolysis (100 mA/cm$^2$) SEM images (Supplementary Fig. 20) show preserved periodic grains and spherical microstructure with sporadic metal clusters as a testimony of strongly adhered highly active and durable catalyst film. Furthermore the excelent structural and activity integrity of the catalyst is confimed from post-electrolysis (100mA/cm$^2$) XPS analysis (Supplementary Fig. 21-22). The Metal dissolution and oxidative corosion is prevalent in Ni-foam compared to CuNiFe particulary

XPS survey spectrum showing a very high O1s peak in Ni-Foam. The elemental peaks show presence of Cu (I & II), Ni (II & III) and Fe (II & III) in CuNiFe (Supplementary Fig. 22).

NiFe has been reported extensively as high performance OER catalysts, however, they suffer long term instability and loss of activity due to insertion of Fe in the bulk of Ni matrix.[52,53] Introducing copper that can enhance Lewis acidity at nickel active centre can increase activity in addition to durability enhancement.[54] To establish the synergism and impact of poly-metal system on alkaline OER we synthesised mon-, bi-, and tri-metallic combinations of Cu, Ni, and Fe metals like Ni, Cu, Fe, CuNi, CuFe, Ni-Fe with the optimised electrodeposition process (5mM of each partipant metal ion, 32 mM $H_2SO_4$, 100mA/$cm^2$ ON pulses for 300 deposition cycles). XRD graphs (Supplementary Fig. 23) shows the deposition of the mono- (Cu, Ni, Fe), Bi- (CuNi, CuFe, NiFe) metallic system. SEM images recorded on mono-,bi-, and tri-metallic (Ni, Cu, Fe, CuFe, NiFe, CuNi) as shown in Supplementary Fig. 24 depict deposition faciltated by addition of copper. Both Ni, and Fe deposits show thin deposition with no discreet particles, whilst Cu deposition shows both particles of spherical and dendritic nature. NiFe shows similar structure to Ni anmd Fe on Ni-foam, that is periodic grains with no identifiable particles. In both CuNi and CuFe particles are easily visible on the repetitve grains on Ni-foam indicating copper rich and faciltated deposition. The samples were tested for OER in 1M KOH in a H-cell to unravel the role of each metal in the ternary system. A noteworthy observation in EIS at open circuit potential (OCP) (Supplementary Fig. 25 a) of mono- and bi-metallic shows charge transfer resistance increasing from mono-metallic (Ni, Fe, or Cu) to bi-metallic system of either combination (CuNi, CuNi, NiFe).

It is widely reported in literature about the synergism in ternary system for OER containing Ni, Fe, and other first row transition elements.[55] The CV polarization curves on different systems (Supplementary Fig. 25b) shows larger kinetic delay (>20 mV on mono-metallic system except Fe (31, 21, 20 mV on Ni, Cu, and Fe respectively) on mono-systems leading to higher overpotentials than bimetallic systems (50 mA/$cm^2$ current reached within <20 mV potential sweep on bi metallic systems (0.19, 0.15,0.19 mV for CuNi, NiFe, CuFe, respectively). The role of increasing ECSA from Ni < Fe < NiFe < Cu < CuFe < CuNi confirms the role of copper is not only ECSA enhancement, but it also modulates the electronic structure in a favourable way to increase the overall OER activity.[41,56,57] NiFe shows least kinetic delay and higher currents only surpassed by Fe at higher potentials, which is expected given the formation of NiFe LDHs and higher peroxidised dissolution (Fe-OOH) of Fe at higher voltages.[58–62] The activity enhancement is often assigned to the large surface area, better intrinsic activity [60], lower activation energy and positive electron modulation often paired with optimum adsorption energy acquired by synergistic tri-metallic system.[63,64] Ni, Fe, Co, Cu based systems occur near the peak in volcano plot for OER in metals and metal oxides[65,66] forming a promising trio for OER.

The synergism between the ternary metal system was further corroborated by analysing mono-, bi-, trimetallic system for steady state OER at different current densities (**Fig. 4a**). **Fig. 4a** shows the overpotential observed on Ni, Cu, Fe, NiFe, CuNi, CuFe, and CuNiFe at various current densities, to probe activity under lower and higher current, 10, 50, 100, 150, 250, 300, 400, and 500 mA/cm$^2$ for one-hour steady state (Supplementary Fig. 26). As expected, Fe exhibits lower overpotentials at 10, 50, and 100 mA/cm$^2$ than Ni, Cu, CuFe and CuNi due to high iron catalysed OER. At higher currents Fe faces higher peroxidised dissolution leading to loss of activity and higher overpotentials. Conversely, presence of other metal increases the stability and activity of the system. CuNiFe shows minimum overpotential among the tested sets at every current density establishing its excellent activity and stability. NiFe[67,68] has been reported as highly active catalysts for OER under alkaline medium. However, the activity at higher currents is not maintained as iron dissolution dominates and Ni-OOH generated degrades. The degradation is due to inclusion of Fe into γ-NiOOH structure and leaching of Fe at higher potentials from this structure leading to system failure.

The presence of a third metal usually reduces such oxidative stress resulting in higher activity.[62] In CuNiFe, copper exists as Cu(0) predominantly (**Fig. 4d**) corroborated from HAADF STEM images, where it prefers core and exists mainly as ternary layer of CuNiFe, CuNi, and CuFeO$_2$ near the interface. The conductive copper core decreases the charge transfer resistance and acts as a lewis basicity site on top to stabilise higher oxidation Ni, and Fe generated during OER. The presence of copper not only reduces the charge transfer resistance but also increases the stability of the system. The presence of a copper-rich basal phase as depicted in TEM (**Fig. 4d**) enhances core conductivity hence reducing the charge transfer resistance. Copper also enhances the surface area due to intrinsic nature of copper depositions yeilding large-surface nanostructures which leads to higher turnover numbers. The role of copper was further probed by synthesising three different CuNiFe from varying concentration of Cu$^{+2}$ (2.5, and 7.5 mM) ions in precursor solution an drest of deposition parameters constant labelled as Cu(2.5 mM)NiFe and Cu(7.5 mM)NiFe to distinguish theconcentration of Cu present. Supplementary Fig. 27 shows the XRD graphs of the deposited samples, both Cu(2.5 mM)NiFe and Cu(7.5 mM)NiFe shows similar cryatlline structures with peaks correspomding to CuNiFe, Ni and Cu.

SEM images shown in Supplementary Fig. 28 depict conformal deposition in Cu(2.5 mM)NiFe sample with islands of polygonic shapes, while as on increasing concentration of Cu to 7.5 mM the formation of dendrites over the particle film is observed. OER in 1M KOH on these samples showed increased activity on increasing Cu concentration (Supplementary Fig. 29). Presence of copper increases activity by taking part in OER activity with further activity enhancement via surface area enhancement. [69] The oxidation peak in Cu(7.5 mM)NiFe has higher area under it compared to Cu(2.5 mM)NiFe demonstrating larger area with active metal sites.[70]

**Discussion**

The surface and bulk properties of an electrocatalyst determine its activity and stability towards OER. [71] To understand the role of microscopic, compositional, and electronic structure on activity and stability of CuNiFe, a combination of transmission electron microscopy (TEM) procedures were used. High resolution TEM (HRTEM) was used for morphological and topological analysis, scanning transmission electronmicroscopy with energy dispersive X-ray (STEM-EDX) aided in deciphering composition of the CuNiFe at different areas. On OER activation for 24 hours where the overpotential for 100 mA/cm$^2$ decreases due to anodic activation[43] the surface reconstruction occurs changing the structure into a layered structure consisting of bi-, and tri-metallic layers unlike traditional layered double hydroxides as observed from STEM-EDX spectrum combined with NMF components analysis (**Fig. 4 b-c**). The top layer of catalyst is restructured into NiFe, CuNiFe, Ni$_3$Fe, CuNi, and CuFeO$_2$ as depicted in FFT pattern derived from HRTEM images at three different sites on the activated catalyst (**Fig. 4d**). Presence of NiFe endows high activity on the CuNiFe, with copper acting as conductive core and facilitating OER by additional generation of OER active sites of CuFe, and CuNi.

As a measure of catalytic OER process, ECSA is an influential factor. ECSA normally measures the intrinsic activity on a catalyst layer.[72] ECSA normalised current is shown in supplementary fig. 16 on CuNiFe. Copper endows higher surface area to CuNiFe which is observed from double layer capacitance of CuNiFe and NiFe (Supplementary Fig. 3) with Cu increasing the ECSA by a factor 7. The oxidation peak preceding OER, which can be attributed to high valance Ni (II, III, and IV) species active during OER, observed on CuNiFe film shows higher current (24.3 mA/cm$^2$) compared to 14 mAcm$^2$ on NiFe (**Fig. 4e**). Presence of copper also facilitates the existence of high valance nickel sites (III and IV) as seen in CV traced at 1mV/s[1] in **Fig. 4e**. The Ni foam shows peak for Ni (0 & II) while as CuNiFe shows presence of higher oxidation states like Ni (III), and Ni (IV). Copper can stabilise the pi-donation from Ni to oxygen therefore aiding in generating highly active oxidation states as noted in literature.[54]

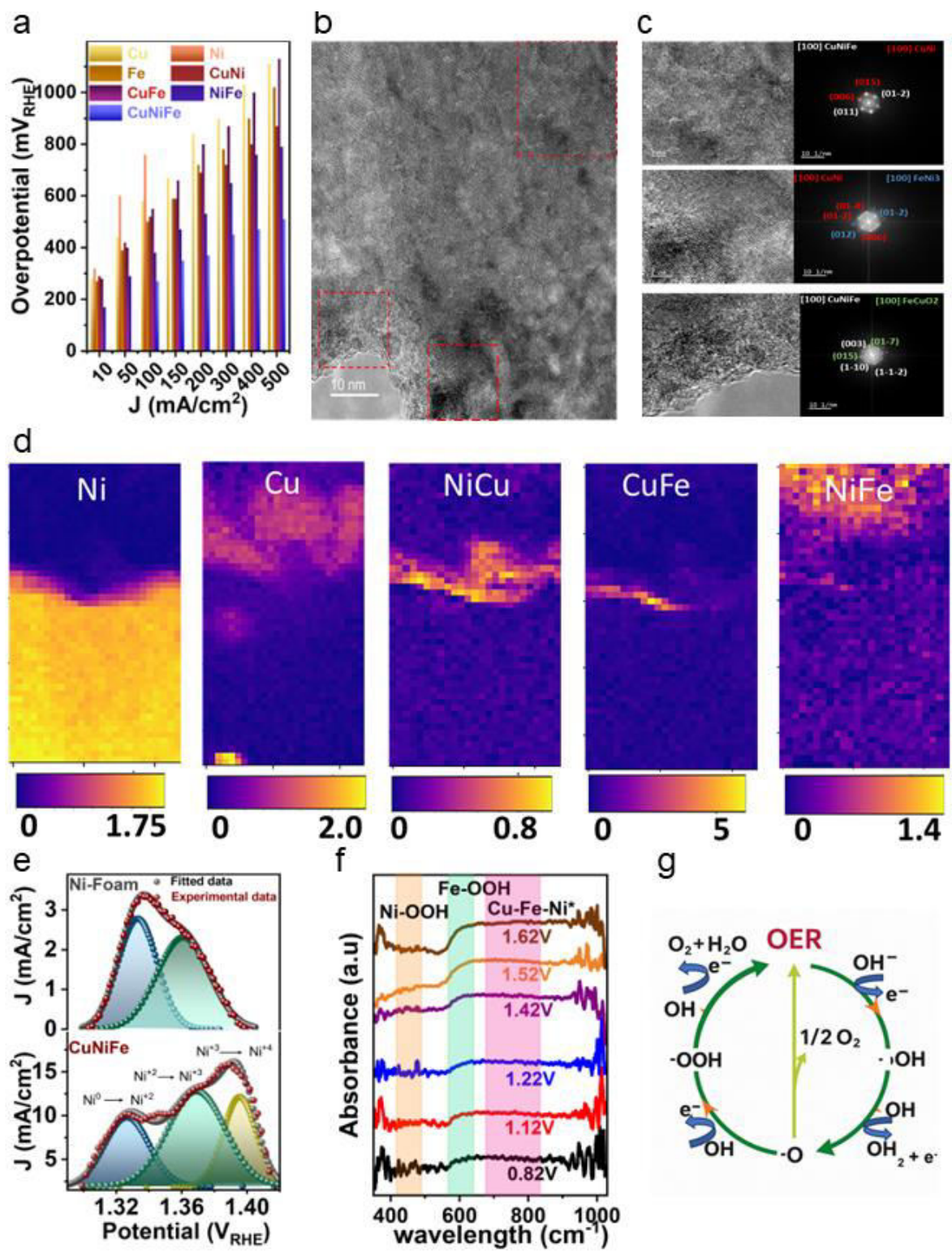

*Fig. 4: Structure-activity relationship. a,* overpotentials at different currents on mono-, bi-, and tri-metallic systems containing Cu, Ni, and Fe, demonstrating synergistic role of metals in OER kinetics. *b-c,* HR-TEM of the activated CuNiFe after 24 hours in 1M KOH and at

*100mA/cm² showing restructuring of the catalyst into CuFeO₂, and Ni-CuFeO₂ phases, the phases are periodic and homogenously on the surface. **d**, elemental distribution of Cu, Ni, Fe after restructuring in 1M KOH. **e**, electrochemical generation of Ni(iv) species during OER on CuNiFe contributing to high rates and lower Fe and Ni peroxide dissolution. **f**, Spectro-electrochemical chrono amperograms on CuNiFe recorded in reflection mode in 1M KOH at different potentials. **h**, mechanism of the OER on the CuNiFe film.*

The OER occurs in two fundamental mechanisms on metal oxides, with different pH dependances.[71,73] The first one is a concerted four proton- four electron transfer pathway, which is pH-independent where activity is mostly as a result of surface metal ion centres.[73] The second one is a non-concerted four proton- four electron transfer which shows dependence on pH for OER activity and is attributed to the surficial and bulk characteristics of the metal centres including lattice oxygen.[71] The activity of CuNiFe anode was tested at different pH (Supplementary Fig. 30a) and indicated a non-concerted proton-electron transfer mechanism operating as an increase in OER activity followed the increase in pH. The variation of overpotentials for 100mA/cm² current is shown in Supplementary Fig. 30b.

The mechanism on CuNiFe was further probed using in-situ UV-Vis spectroscopy. **Fig. 4f** shows the evolution of Ni-OOH, Fe-OOH and Cu-Ni-Fe-O[74–76] active sites with increasing potentials. At 0.82 V there is already a slight faradaic signature due to Fe-OOH production. With increased applied potential the peaks for Ni-OOH, Fe-OOH (also A2g to T2g transition of Ni (II)) progressively increase as well. The peak for CuNiFe occurs from 700-800 nm, suggesting synergistic role of the ternary system. The similar studies on Ni foam (Supplementary Fig. 31) shows no peak emergence at till 1.42 V when a peak for Ni-OOH (400-450 nm) and Fe-OOH (580-650 nm) starts emerging. There are no broad peaks around 650-800 nm on Ni-foam which are present on CuNiFe suggesting ternary -OOH active system, potentially Cu-FeNi-OOH as supported by TEM studies. As confirmed from pH dependence of activity, active site determination and absorbate pattern through spectroelectrochemistry the mechanism operable is non-concerted as shown in **Fig 4g.**

 **Full-cell alkaline water electrolyser**

CuNiFe was further tested for water electrolysis in a two electrode configuration in H-cell in 1 M KOH and 25°C. Supplementary Fig. 31 shows the cell potentials at –100 mA/cm² at standard ambeint tempertaure and pressure (SATP) conditions for different electrode couples. For IrO₂‖Pt couple the cell potential is 4.5 V and increases on average by 10 mV/h, which is attributed to sluggish alkaline hydrogen evolution on Pt.[77] However, Pt is very stable under alkaline conditions and was chosen to compare the effect of CuNiFe as an anode on the overall activity and stability of the full water electrolysis. The activity increased going from IrO (4.5 V) < Ni-foam (4.05 V) < CuNiFe (3.5 V). It was also shown that CuNiFe can work as hydrogen evolution cathode in alkaline solutions with activity slightly better than Pt as cathode and better durability for more than 100 hours.

We then explored the practical utility of CuNiFe as an anode for alkaline exchange water electrolysis (AEM-WE) in zero-gap configuration (Supplementary Fig. 32). In AEM-WE assembly Sustainion™ was used as ion exchange membrane, nickel foam as cathode and 30% KOH as flowing electrolyte. **Fig. 5a** shows the CVs (iR corrected and uncorrected) recorded at 25 °C on CuNiFe in 30% KOH. At 1.73 V, a 1 A/cm$^2$ current density was measured. Galvanostatic EIS at different currents (10, 100, and 1000 mA/cm$^2$) were recorded to probe the total resistance of the electrolyser as shown in **Fig. 5b**. At 1000 mA/cm$^2$ currents, the total cell resitance dropped toless than 1 ohm showing a superior activity at such large gas production currents. EIS at different potentials are shown in Supplementary Fig. 33. The total charge transfer resistance at thermoneutral potential (1.48) is ~10 Ω, showing superior catalytic activity of CuNiFe. The catalyst was tested in these low temperature industrial conditions for 500 hours. **Fig. 5c** shows the chropotentiometric graph recorded on CuNiFe. The cell voltage is 1.85V at 1 A/cm$^2$ showing one of the best performance at such drastic conditions. A voltage efficiency of 66% is achieved at 25 °C, 1A/cm$^2$, and 30% KOH.

We further validated the catalyst for alkaline water electrolysis (AWE). In a set-up similar to Supplementary Fig. 32, Zirfon™ was used as diaphram to separate anode and cathode. **Fig. 5d** shows the CV polarization curve measured on CuNiFe anode paired with nickel foam cathode.

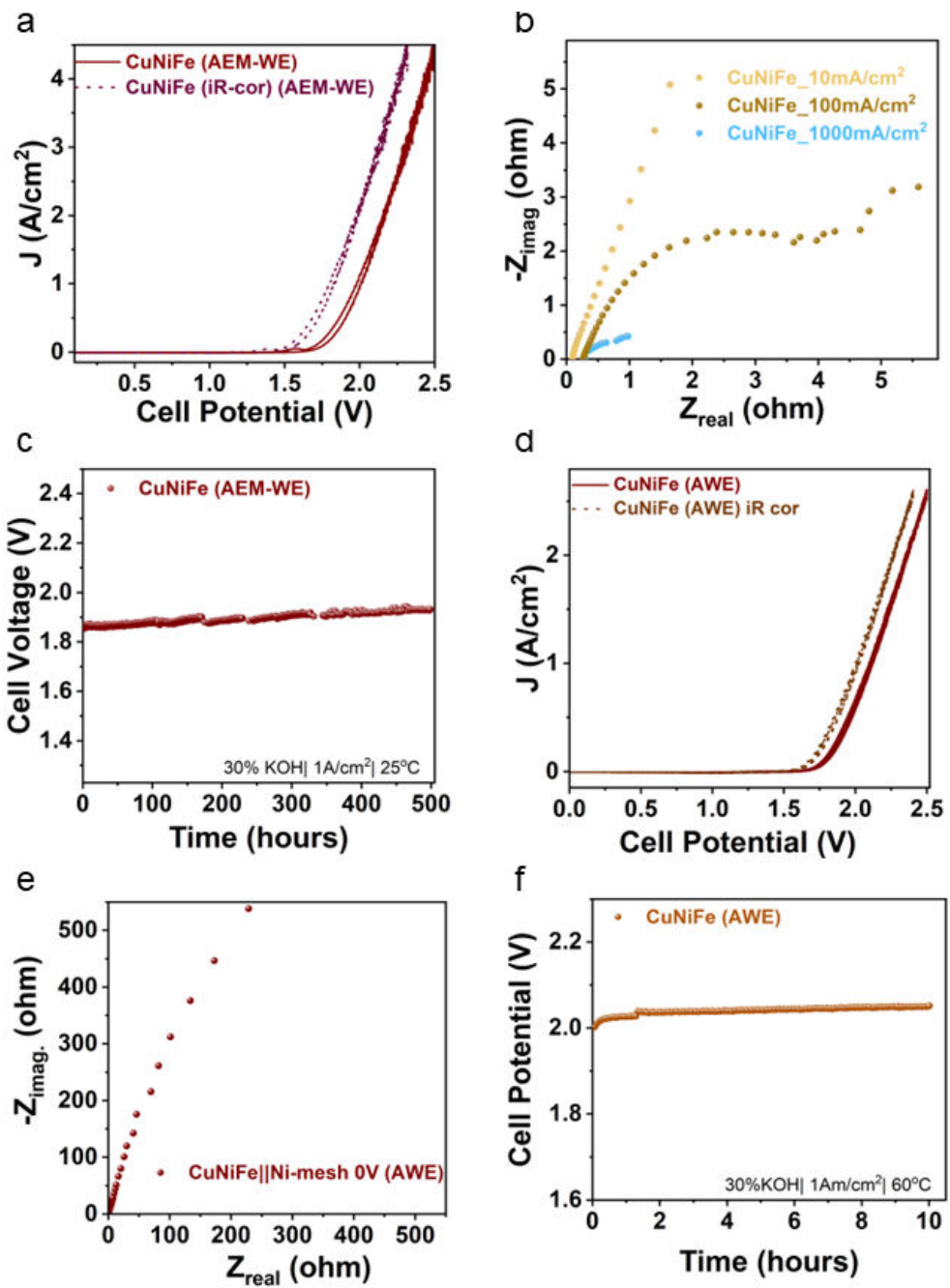

**Fig. 5: Practical application for AWE and AEM-WE at industrial conditions.** *a, CV (iR corrected and uncorrected) obtained in full cell configuration in AEM-WE. b, Galvanostatic EIS traces obtained on CuNiFe in AEM-WE configuration at 10, 100, and 1000 mA/cm² currents. c, Stability of the CuNiFe film for AEM-WE at 25ºC for 500 hours. d, CV traces of CuNiFe in a AWE configuration at 60ºC. e, Potentiostatic EIS at 0V on CuNiFE in a AWE configuration. and f, Chrono potentiometric traces of the CuNiFe at 1 A/cm² and 60 ºC temperature in AWE configuration.*

At 2 V, a 1 A/cm² current density is obtained at 60ºC. The EIS at 0 V is shown in **Fig. 5e**. The overall cell resistance is higher as diaphrgm does not allow direct ionic tranfer as observed in AEM. **Fig. 5f** shows the water electrolysis at 1 A/cm$^2$ in 30% KOH at 60 °C for 10 hours. There is no degradation confirming the practical use of CuNiFe in both AWE and AEM-WE systems. Supplementary Fig. 34 shows the potentials at different applied currents in AWE configuration. The CuNiFe film was stable for the testing time of 6 hours at 2.5 A/cm$^2$ with a potential of 2.5 V (iR uncorrected). This shows the synthesised CuNiFe film holds great potential to extend operation window of commercial AWE.

**CO$_2$ electrolysis**

The role of anodes in alkaline electrolysis has been mostly studied in water electrolyser.[78] However, there is also a strong interest in elucidating the role of anode composition in other types of alkaline electrochemical processes, such as the electrochemical reduction of carbon dioxide (CO2RR).[79] Anodes in CO$_2$ electrolysers are challenging to optimize given the mass transfer from cathode side to anode side and reoxidation of products or carbonate into CO$_2$ often resulting in salt deposition on the anode, leading to its failure. An efficient NiFe anode paired with cobalt thiocyanine-carbon nanotube composite cathode showed >97% CO FE at ~3 V cell potential.[2]

To ascertain the effects of CuNiFe anode on the faradaic efficiencies and current density from CO$_2$ reduction, CuNiFe was paired as an anode with Ag nanoparticles on GDE in 25 cm$^2$ zero-gap cell. The flow of CO$_2$ was 50 ml/min in 0.5 M KHCO$_3$ at 25°C. **Fig. 6a** show the FEs on Ag-GDE with Ni foam and CuNiFe as anode. On pairing with CuNiFe the current density increased by 0.2 mA/cm$^2$ at 2 V to 30 mA/cm$^2$ at 3 V. A decrease in HER was observed with production of CO with 96.3% FE, and partial current density of 26.4 mA/cm$^2$ at 2.5 V, which is an increase of ~8% in FE and 9.2 mA/cm$^2$ in partial current density for CO on Ni foam as anode at similar potential. Furthermore, a similar experiment paired CuNiFe anode with commercial 40% Cu-Vulcan gas diffusion electrode (Cu/C GDE) cathode in 1M KOH for CO$_2$ reduction **(Fig. 5b)** in 25 cm$^2$ zero-gap cell. The CO$_2$ flow rate was 100 ml/min with an electrolyte flow rate of 5 ml/min. An interesting finding is that CuNiFe as anode triples the current density in CV curves (Supplementary Fig. 35) compared to Ni foam and enhances the FEs to CO$_2$ conversion products.

Interestingly, at lower voltages (2.0 V, 2.25 V, 2.5 V), CuNiFe as anode: exhibited a decrease in HER along with formation of C2 products (like ethanol and ethylene) in higher fraction than C1 products, unlike observed for Ni foam as anode on the same cathode. These results show that the fabricated electrode is not only an excellent water electrolysis anode but also a promising CO$_2$ electrolysis anode. The increased activity can be attributed to the fascile half oxidation reaction allowing higher cathodic potential at same applied voltage. Further investigations are underway to understand the role of anode in steering the CO$_2$ conversion products in CO$_2$ electrolysis, which currently has not been reported or expounded and can open advanced ways to promote the CO$_2$ reduction reactions.

**Scalability of anode**

One of the challenges of translating laboratory designed electrodes to industrial use is the scale up of the electrodes. Most of the electrodes fabricated, that show promising activity are often difficult to scale up, expensive or loose material homogeneity and electrochemical characteristics on scale up. We used the optimised electrodeposition method (5 mins, 25 °C) to deposit CuNiFe anode of 50 cm$^2$ to test the process' scalability and retention of the desired physical, chemical, and electrochemical properties. **Fig. 6c** shows the digital image of the large electrode deposited on Ni foam. From the macro image (Supplementary Fig. 36) the surface of the large electrode is smooth with full coverage of the foam exposed to the electrolyte without any loose aggregates or powder.

To ascertain the homogeneity and uniform deposition, 5 pieces of 1 cm$^2$ were extracted from large electrode at random places. The extracted pieces were subjected to XRD, SEM analysis to check the crystalline and microscopic homogeneity. XRD of these five extracted points from the large electrode selected randomly and labelled as CuNiFe-A1, A2, A3, A4, and A5 shows similar pattern to XRD of 1 cm$^2$ deposited CuNiFe (Supplementary Fig. 37). The SEM images (Supplementary Fig. 38) show same topological and morphological uniformity on all points of extraction confirming the homogenous deposition throughout the large electrode, which is one of the biggest challenges of scaling up. After confirming the retention of physical properties by large electrode, we also tested the electrochemical properties of the five extracted electrodes for their OER activity (Supplementary Fig. 39) to evaluate the electrochemical scalability of the surface.

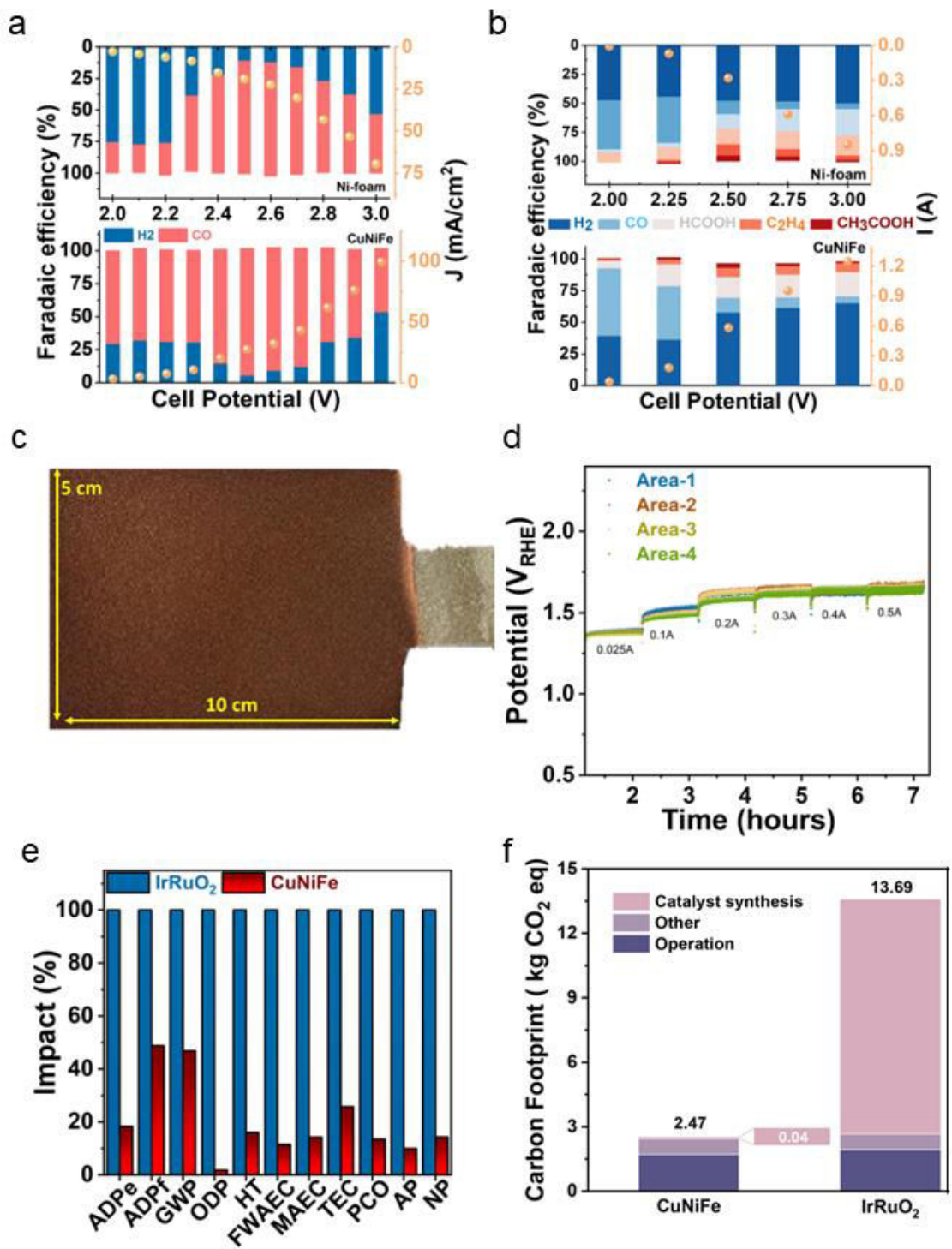

**Figure 6: Practical application towards CO₂RR, scalability and sustainability. *a,*** *current densities and Faradaic efficiencies at different cell voltages for syngas using Ag NPs on GDE as cathode in 0.5 M KHCO₃ and Ni foam and CuNiFe as anodes, both FE and current densities are enhanced with CuNiFe anode. **b,** product distribution on Cu/C GDE cathode in MEA and 1M KOH obtained on Ni foam and CuNiFe as anodes, a triple-fold increase in current density and carbon rich products are obtained in higher fraction on CuNiFe anode. **c,** digital image of 50 cm² electrodeposited film. **d,** currents at 5 extracted pieces from the large electrode showing*

*similar activity. **e,** comparison of impact of CuNiFe on environment with IrRuO$_2$. **f,** carbon footprint of CuNiFe compared to IrRuO$_2$.*

**Fig. 6d** shows chronopotentiometric graphs obtained on these five electrodes. The electrodes behave similarly at different current densities from 25 to 500 mA/cm$^2$. The results are very interesting from industrial translation point of view. The large electrode of 45 cm$^2$ area was tested in 1 M KOH at 25 °C for its OER activity in a two-chamber cell. Supplementary Fig. 39 a shows CV traces on the extracted pieces with similar currents. Furthermore, Supplementary Fig. 39b shows the cell potential for 4.5 A current on 45 cm$^2$ electrodes. The potential is similar (1.51 V) to what was observed while using 1 cm$^2$ (1.57 V) which confirms that the method can be used to fabricate large scale electrodes without compromising on physical, chemical, or electrochemical properties of the CuNiFe catalyst.

Environmental sustainability is becoming increasingly critical in catalyst development, particularly for electrochemical processes such as the OER, which plays a crucial role in renewable energy systems, including water electrolysis for hydrogen evolution reaction (HER) and carbon dioxide reduction (CO$_2$RR) to organic chemicals. Currently, Iridium and Ruthenium oxide derivates catalysts (IrRuO$_2$) remains the benchmark catalyst due to their exceptional catalytic activity and stability under harsh reaction conditions.[80] However, concerns related to high cost, scarcity of iridium resources, and significant environmental impacts associated with its extraction and processing have intensified the search for alternative catalysts.[81,82]

We conducted a comparative Life Cycle Assessment (LCA) between newly synthesized CuNiFe and conventional IrRuO$_2$ catalysts to evaluate their environmental footprints comprehensively across their respective life cycles, aiming to identify a more sustainable catalytic solution for practical OER applications. In this study, the LCA followed ISO 14040/14044 standards,[83] defining a functional unit of 1 kg of oxygen production in Anion exchange membrane (AEM) cell. A cradle-to-gate system boundary was defined in this work, as shown in Supplementary Fig. 40 includes catalyst synthesis, electrode preparation, membrane usage, electricity and electrolyte consumption. **Fig. 6e** shows the environmental impact of CuNiFe as an anode compared to IrRuO$_2$ or PGM based anodes. CuNiFe has 40-60% more positive environmental impacts across all impact categories thereby confirming the sustainable nature of the developed anode. Since the main constituent is copper so its recyclability is also very high. **Fig. 6f** shows the total carbon footprint emerging from using PGM/IrRuO$_2$ anodes compared to CuNiFe which is minimum 10 times lesser. The global warming potential of using IrRuO$_2$ is almost triple than CuNiFe, which is due to expensive and complex supply chain.

**Conclusion**

In summary, we have developed a scalable, room-temperature electrodeposition process to fabricate a novel CuNiFe anode that demonstrates exceptional activity and durability for the oxygen evolution reaction at industrial-scale current densities. The catalyst's unique core-shell electronic structure, with a conductive Cu core and a Ni-rich active surface, enables low overpotentials and remarkable stability exceeding 500 hours in harsh alkaline conditions. Its efficacy is proven not only in advanced water electrolysers (AEM-WE and AWE), achieving high current densities at low cell voltages, but also in enhancing the performance of CO$_2$ reduction systems by boosting current densities and shifting product selectivity. Combined with a significantly lower environmental footprint compared to noble-metal benchmarks, the CuNiFe anode represents a critical step towards economically viable and sustainable electrochemical technologies for green hydrogen and chemical production.

**Methods**

**Chemicals**

The nickel foam (95 porosity, 99.9%, 0.9 mm thickness) and mesh (0.2 mm thick, 99%) was procured from Goodfellow UK. H$_2$SO$_4$ (96-98% purity), KOH (Fe free, 99.99%) were purchased from Fischer scientific. Metal precursors, nickel nitrate hexahydrate (99.9%), copper chloride dihydrate (99.9%), and iron chloride hexahydrate (99.9%) were purchased from Merck. Platinum meshes were procured from AV supplies China. IrO$_2$ gas diffusion layer (GDE), sustainion anion exchange membrane (RT- 37X) from Dioxide materials USA.

**Catalyst preparation**

The synthesis of CuNiFe was carried out by electrodeposition. In a conventional 3-electrode cell, at room conditions (25 ºC), 5mM each of CuCl$_2$.2H$_2$O, NiNO$_3$.6H$_2$O, and FeCl$_3$.6H$_2$O aqueous solutions were added. The deposition substrate was Titanium foil (Ti) or nickel foam (Ni-Foam)[84] with platinum[85] as counter electrode and gel Ag/AgCl as reference electrode. Electrolytic bath was kept at acidic pH for an optimised deposition competition between the metal ions using H$_2$SO$_4$ (20-40 mM). The ternary metal system was deposited using unipolar pulses of -100 mA/cm$^2$ amplitude for 0.5 s followed by null pulses for 0.05-0.5 s for 100-300 cycles. The null pulses were timed to depolarize the system for a homogenous repopulation of metal ions around the deposition substrate (supplementary figure 1). Nickel foam used was pretreated with acetone, 3M HCl (sonication for 10 minutes, 25 °C), and water in that order to get rid of organic carbon impurities and any nickel oxides. Different number of pulse cycles (100-350) were used to optimise the catalyst loading and surficial properties. CuNiFe was also optimised for pH of electrolytic bath, and deposition surface for desired nanostructures and compact layer. Ni mesh was pretreated similar in way to nickel foam.

Cathode preparation for $CO_2$ reduction: The cathode electrocatalysts were processed as an ink by suspending 10 mg of the material powder in 2 mL of absolute ethanol and adding 10 μL of 5v% Nafion solution in ethanol/water. The ink was sonicated for 20 min in an ice bath and spray-coated on a 5 cm$^2$ square piece of carbon-based gas-diffusion layer (AvCarb GDS5130) with a target loading of 1 g/cm². The electrode was then dried for five hours at 80°C.

**Catalyst characterization**

X-ray diffractograms were obtained on Malvern Pananlytical Aeris with Cu source in thin film setting. SEM images and EDS maps were captured with Zeiss EVO LS15. High-angle annular dark field scanning transmission electron microscopy (HAADF STEM) imaging and compositional mapping using energy-dispersive X-ray spectroscopy (STEM-EDX) were performed using a FEI Tecnai Osiris FEG-(S)TEM operated at 200 kV. In this experiment, a standard STEM probe with a beam current of 140 pA (70 μm C2 aperture, spot size 6, and gun lens 5) was used. HAADF STEM images were acquired with a Fischione detector, with a spatial sampling of 4 nm per pixel and a dwell time of 0.5 s. EDX spectra were obtained with Bruker Super-X SDDs with a total collection solid angle of ≈0.9 sr, a spatial sampling of 4 nm per pixel, a dwell time of 50 ms, and a spectral resolution of 5 eV per channel. For EDX the electron dose is estimated as <7,000 e⁻Å$^{-2}$s$^{-1}$. All EDX spectra were spectrally rebinned to 10 eV per channel, then denoised with PCA/NMF and processed in HyperSpy 1.7.3. Atomic resolution TEM images are acquired from a FEI Tecnai Osiris FEG-(S)TEM operated at 200 kV, with a spot size of 9 with a beam current of 60 pA. The image was then denoised with HyperSpy 1.7.3 with Gaussian fitting. The TEM lamella cross-section was prepared using an FEI Helios Nanolab Dualbeam FIB/SEM following standard protocols, down to 100nm to be electron transparent. The lamella was transferred with minimal air exposure into a FEI Tecnai Osiris FEG-(S)TEM.

Raman spectra were collected with Renishaw inVia™ (confocal, 532nm laser). X-ray photoelectron spectroscopy (XPS) data were collected on a Thermo Scientific K-alpha photoelectron spectrometer with a monochromatic Al Kα radiation (photon energy of 1486.6 eV), a hemispherical analyser and two–dimensional detector. The electron energy analyser consists of a double focusing 180° hemisphere with a mean radius 125 mm, operated in constant analyser energy (CAE) mode, and a 128-channel position sensitive detector. Measurements were conducted with a 400 μm spot size, using a dual beam flood gun (electron and Ar+ ion) with 100 mA current. Survey and core level spectra were collected at pass energies of 200 eV and 50 eV, respectively. The elemental quantification based on the collected core level spectra was performed using the Thermo Scientific Avantage Data System software package (v5.9925). A Shirley background was applied for all spectral fitting. ICP-MS (Agilent 7900, single-quad) was used to determine Ni, Cu, Fe ions in the solutions after filtration for any insoluble particles. In-situ UV-visible spectroscopy (Uv-Vis) was done with Spellec from Metrohm in reflection mode. The cell was operated under reflection mode. In a typical experiment a horizontal electrode was connected to the potentiostat, and reflection probe was placed on it with a distance ~1mm from electrode surface. A 10 ms data collection was used per data point collection in a 10 minutes chronoamperometry analysis.

**Electrochemical measurements**

Electrochemical measurements were recorded in 1M KOH in a 3-electrode H-cell unless stated otherwise with Hg/HgO reference electrode, and Pt as counter electrode on Vionic potentiostats from Metrohm. The pH dependence studies were carries in different concentrations of KOH solution (0.05M, 0.1M, 0.2M, 0.5M, 1M, 6.89M). The temperature studies were done in H-cell in different KOH solutions. In a zero-gap cell, 1M KOH was used at $60^0C$ to test the performance of CuNiFe. Selected potentials and overpotentials are reported with maximum 30% iR correction. All measurements are reported with respect to reversible hydrogen electrode using following potential conversion formula:

$$E_{RHE} = E_{Hg/HgO} + 0.098V + (0.059 * pH)$$

Where $E_{RHE}$ is the potential with respect to reversible hydrogen electrode (RHE), $E_{Hg/HgO}$ is the potential observed with Hg/HgO reference electrode, 0.098V is the potential difference between RHE and Hg/HgO at 25 °C in 3M KOH, and pH is the pH of solution.

All electrochemical measurements unless stated otherwise are iR uncompensated for best possible activity in H-cell configuration. The zero-gap cell was assembled with CuNiFe as anode pressed on sustainion anion exchange membrane, and Nickel foam as cathode. The electrolyte flow rate was 10 ml/min with temperature control through hot plate with temperature sensor. For electrochemical active area (ECSA), electrochemical double layer capacitance ($C_{dl}$) was obtained by CV (0.0-0.35V and back) Vs Hg/HgO under various scan rates (20-400 mV/s). To calculate the ECSA, 60μF capacitance was used as reference recommended for porous substrates by Bard[86].

**Product measurements**

For alkaline water electrolyser GC (Agilent, 8890) was used to determine the faradaic efficiency (FE) and molar production of hydrogen and oxygen from water splitting.

$FE = Q_{output}/Q_{input}$

$CO_2$RR experiments were performed inside an electrochemical flowcell provided by Dioxide Materials and composed of two stainless steel flowplates and a membrane-electrode assembly. The membrane-electrode assembly was composed of a 5 cm² cathode and a 6.2 cm² anode (Ni foam or CuNiFe on Ni foam, cut with scissors to the appropriate shape) sandwiching a Sustainion® X37-50 membrane. The cathode side was supplied with a 50 mL/min stream of pure $CO_2$, humidified by bubbling through a water reservoir prior to entering the cell, while the anode side was supplied with a recirculating 5 mL/min stream of 0.5 M $KHCO_3$ (Ag cathode) or 1 M KOH (Cu/C cathode), forming a closed loop with a 30 mL reservoir. Cyclic voltammetry and chronoamperometry were performed by means of an OGF01A potentiostat provided by Origalys. Product detection was performed during and at the end of 30 min chronoamperometries by in-line micro-GC (Inficon micro-gc Fusion) for gas products, and by NMR for liquid products (an aliquot of 0.1 mL was mixed with 0.3 mL of $D_2O$ to prepare each NMR sample). Faradaic efficiencies were calculated by considering the amount of product generated and the total amount of product that can be generated at applied currents.

**Acknowledgements**

The authors gratefully acknowledge funding received from the European Union's Horizon Europe research and innovation program for the SUNPEROM project, Grant Agreement No. 101223212. M. Abdi-Jalebi acknowledges University College London's Research, Innovation and Global Engagement, UCL – Korea University Strategic Partner Fund for their financial support. The authors wish to acknowledge the support of the Henry Royce Institute for Advanced Materials through the Industrial Collaboration Programme (RICP-R4-100061) and MATcelerateZero (MATZ0), funded from a grant provided by the Engineering and Physical Sciences Research Council EP/X527257/1. The authors acknowledge the Department for Energy Security and Net Zero (Project ID: NEXTCCUS), University College London's Research, Innovation and Global Engagement, University of Sydney – University College London Partnership Collaboration Awards, UCL-Peking University Strategic Partner Funds, Cornell-UCL Global Strategic Collaboration Awards and IISc-UCL Joint seed fund for their financial support. The authors acknowledge the ACT program (Accelerating CCS Technologies, Horizon2020 Project No. 691712) for the financial support of the NEXTCCUS project (project ID: 327327). This work was supported by the Henry Royce Institute for advanced materials through the Equipment Access Scheme enabling access to the Royce SEM-FIB Suite at Cambridge; Cambridge Royce facilities grant EP/P024947/1 and Sir Henry Royce Institute – recurrent grant EP/R00661X/1.

**Author information**

M. Abdi-Jalebi supervised the project. N. Rashid and M. Abdi-Jalebi designed, conceived the key ideas and carried out electrochemical investigations. S. Yang, Z. I. Albu, G. Sanfo and I. Erwig carried some of repeatability and pH electrochemical investigations. P. Bhat and R. G. Palgrave did the XPS analysis. X. Li, T. Wu and C. Ducati performed TEM image analysis including HAADF imaging, elemental distribution and NMF factoring. M. Zendehdel provided input on manuscript and tested some anodes for OER. L. Piccolo and M. Prevot performed the $CO_2$ reduction reactions and product determination.

**Data availability**

All data needed to evaluate the conclusions in the paper are present in the paper and/or Supplementary Information. Source data can be provided with this paper. Additional data related to this paper may be requested from the authors.

# Boosting high-current alkaline water electrolysis and carbon dioxide reduction with novel CuNiFe-based anodes

Nusrat Rashid[a], Shurui Yang[a], Galyam Sanfo[a], Isabelle Ewing[a], Zahra Ibrahim Albu[a], Xinjuan Li[b], Tianhao Wu[b], Prajna Bhatt[c,d], Mathieu Prevot[e], Laurent Piccolo[e], Mahmoud Zendehdel[f], Robert G. Palgrave[c], Caterina Ducati[b], Mojtaba Abdi-Jalebi[a*]

[a] *Institute for Materials Discovery, University College London, London, WC1E7JE, UK*

[b] *Department of materials Science and Metallurgy, University of Cambridge, Cambridge, CB30FS, UK.*

[c] *Department of Chemistry, University College London, London, WC1H0AJ, UK.*

[d] *Istituto Officina dei Materiali (IOM)-CNR, Laboratorio TASC, in Area Science Park, S.S.14, Km 163.5, Trieste I-34149, Italy*

[e] *Université Claude Bernard Lyon 1, CNRS, IRCELYON, Villeurbanne, F-69100, France.*

[f] *Iritaly Trading Company S.R.l., Via Volturno 58, 00185 Rome, Italy*

[*]*Email: m.jalebi@ucl.ac.uk*

## Table of Contents

**Synthesis and characterisation of CuNiFe**

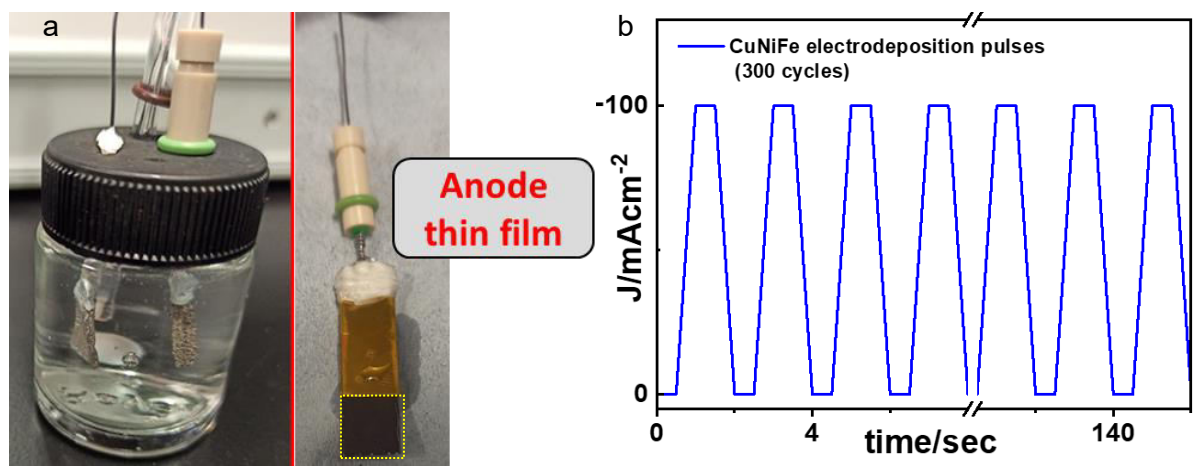

**Supplementary Fig. 1:** (a) Digital image of 3-electrode cell used to deposit the 1-4 cm$^2$ electrodes and a freshly deposited electrode on the left (yellow square showing the thin film), and (b) optimised pulsed program for electrodeposition (300 cycles with each cycle consisting of ON pulse (-100mA/cm$^2$) of 0.5s and OFF pulse (0mA/cm$^2$) for 0.05-0.5s) of the thin CuNiFe film on Ni-foam substrate.

**Supplementary note 1: Substrate optimisation**

300 cycles of ON (-100 mA/cm$^2$) and OFF (0 mA/cm$^2$) current were used to deposit CuNiFe from 5mM each of Cu, Ni, and Fe aqueous bath on titanium foil (CuNiFe@Ti) and porous nickel foam (CuNiFe@Ni Foam). Supplementary fig 2 (a) shows the CV polarization curves of CuNiFe film on titanium foil and nickel foam. Under lower overpotentials the activity on Ti foil as well as Ni-foam supported CuNiFe is almost similar and then it increases more in Ni foam substrate compared to Ti substrate which can be alluded to better gas escape and larger surface area. This observation is a confirmation towards the intrinsic nature of the activity of the deposited metal system rather than a mere surface area dependent activity enhancement. The intrinsic nature of activity is further proved by testing durability of the system deposited on Ti foil and Ni-foam by chronopotentiometry at 100mA/cm$^2$ as shown in Supplementary fig 2 b. Both catalysts show stable operation for more than 24 hours with Ni foam supported CuNiFe operating at least 100mV less overpotential for similar currents. The bare titanium foil forms titanium oxide and hydroxide film which has very low conductivity in KOH solution leading to no activity.[1] However, coating with the CuNiFe film made the system work well, indicating a homogenous conformal thin film deposition.[2] The intrinsic nature of the activity enhancement is further corroborated by electrochemical active surface analysis of Ni-foam and CuNiFe@Ni-foam as shown in Supplementary fig 3. There is a very low enhancement in diffusion layer capacitance from Ni foam to CuNiFe, (from 10.2 to 11.1 mF/cm$^2$) suggesting the enhanced activity is not merely a manifestation of increased surface area however the very intrinsic nature of the activity stemming from well adhered thin film with unique microscopical and electronic structure (Supplementary Fig. 4). The composition confirmation from XPS (Supplementary Fig. 4) shows presence of Cu, Ni and a successful incorporation of small amounts of Fe.

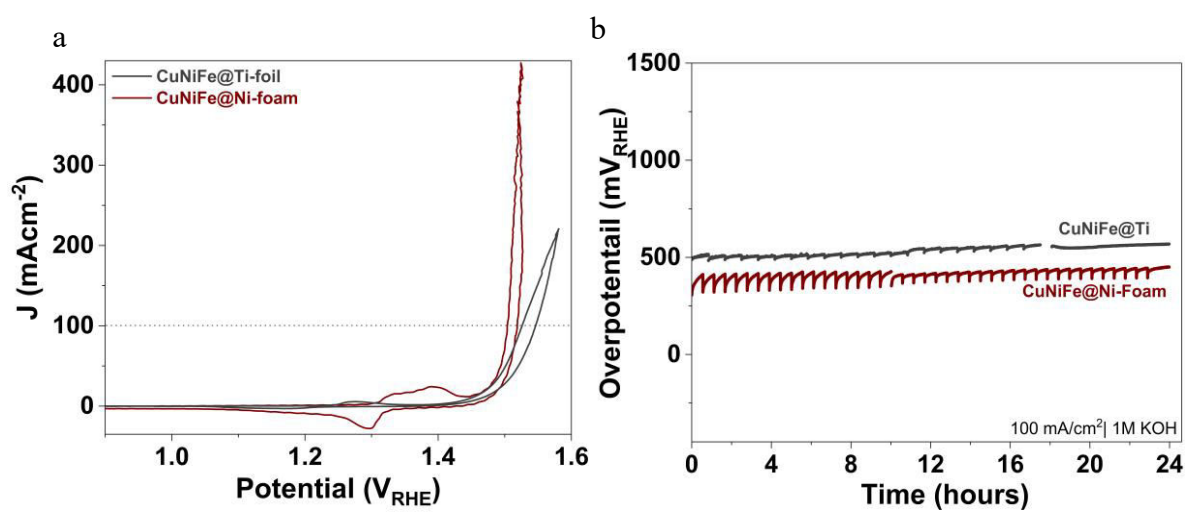

**Supplementary Fig. 2:** Effect of substrate surface area and porosity on OER activity of CuNiFe at lower and higher currents. (a) CV traces of deposited CuNiFe on solid titanium foil and nickel foam, the CVs are 80% iR corrected and shows almost similar activity at lower potentials while deviating at higher potentials due to advantageous gas escape and larger surface area available in Ni-foam[3] (b) chronopotentiometric stability curves on CuNiFe@Ti, and CuNiFe@Ni-foam showing nickel foam as a better substrate at higher currents (>100 mA/cm$^2$) with lower overpotentials (~100 mV lower).

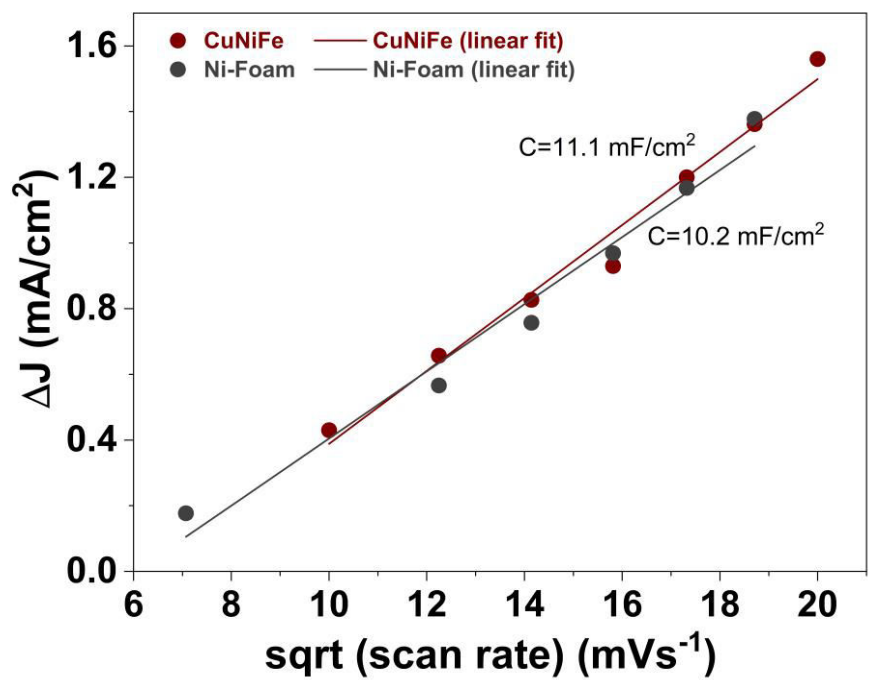

**Supplementary Fig. 3:** Double layer capacitance of CuNiFe and Ni foam calculated from CV recorded in non-faradaic region for both CuNiFe and Ni foam. There is only a small increase (from 10.2 to 11.1 mF/cm$^2$) in surface area on deposition of CuNiFe on Ni-foam confirming conformal layer deposition.

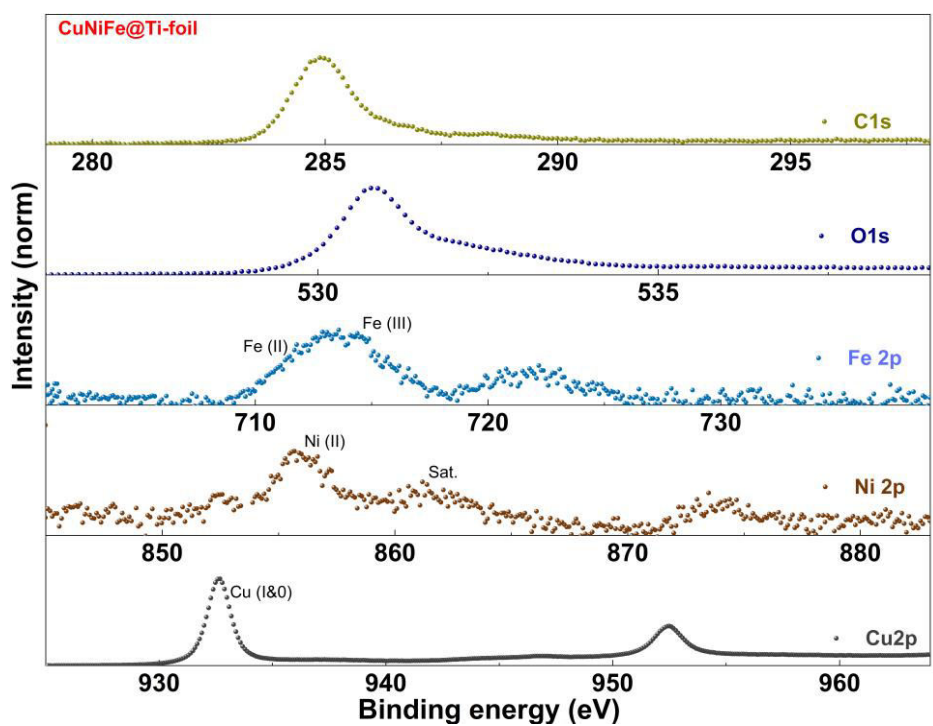

**Supplementary Fig. 4:** XPS obtained for Cu, Ni, Fe, C and O on CuNiFe@Ti-foil. The graphs show presence of Cu (0 & I), Ni (II), and Fe (II, III), oxidation states, -OH metal bonds from O1s and adventitious carbon in C1s.

**Supplementary note 2: Optimisation of pH of electrolytic bath**

The pH of solution has a tremendous effect on the nature of metal deposition and grain structure through regulation of limiting current density.[4] Metals like iron and nickel are very susceptible to pH changes[5] because of their high electrochemical polarization potential, this leads to controlled deposition content in a mixed bath of class II metals like copper. In presence of Fe (III) copper deposition is restricted [6,7] and together with electrode potential of Cu>Ni>Fe leads to a controlled homogenous small grain size. The competition between Fe, Cu and Ni with an acidic pH ensures impurity level of iron deposition in copper rich deposits. To optimise the pH for CuNiFe ternary metal deposition a series of concentrations of $H_2SO_4$ were added while keeping equimolar concentration of Cu, Ni, Fe. The materials tested for alkaline water oxidation shows different activities because of different morphologies and film structure as shown in Supplementary fig. 5.

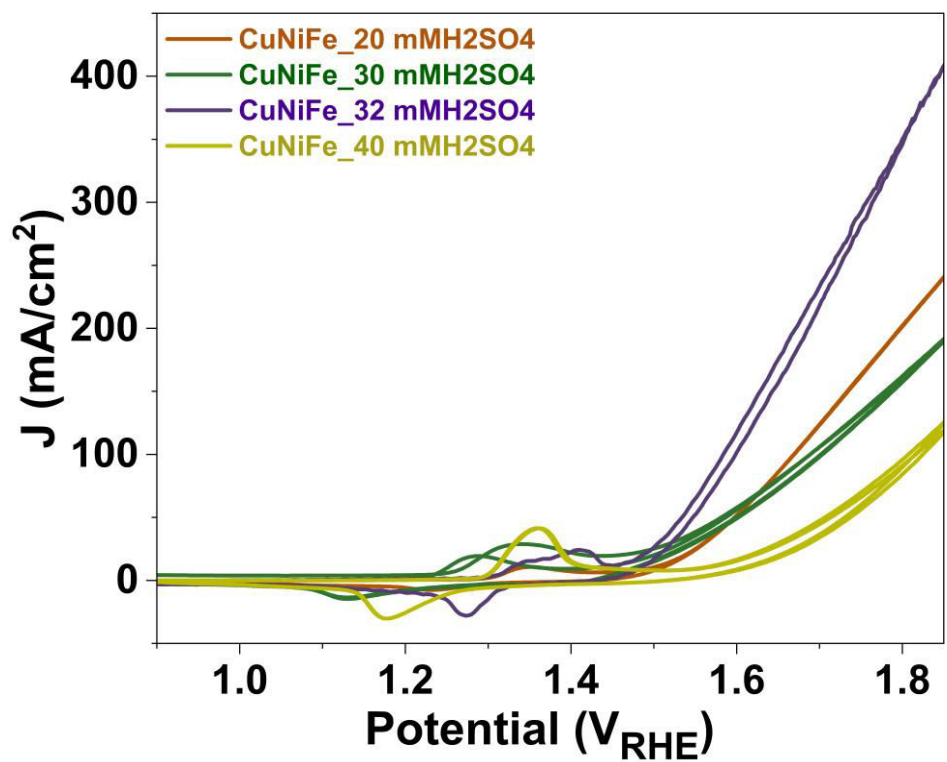

**Supplementary Fig. 5:** CV polarization curves of CuNiFe deposited from similar electrolyte baths (containing equimolar (5 mM) of Ni (II), Fe (III), and Cu (II) ions) at varying pH steered via different concentration of added $H_2SO_4$ solution in 1M KOH in a H-Cell.

**Supplementary note 3: Optimisation of deposition time**

The pulsing current was chosen such that the co-deposition of Ni and Fe is possible along with Cu being well above the standard potential requirements of each metal. However, the pulsing frequency and pH optimisation allows the co-deposition with controlled concentration of the bath metals. The thickness of deposited film is dependent on current efficiency and the time of electrodeposition.[8] A prolonged electrodeposition at higher current efficiencies lead to dendritic or asymmetrical larger dimensional structures, with irregular growth and a thick deposition.[9,10] There is an interplay between optimal mass loading of a material with its electrochemical activity. Oftentimes smaller loading leaves much of the substrate exposed leading to lesser activity and possible direct substrate participation and a very thick film which shows activity like bulk. We optimised the electrodeposition time so that maximum activity is obtained from deposited samples. The electrodeposition method involved same parameters as in main recipe except changing the number of pulses a)100 cycles, b) 200 cycles, c) 300 cycles, and d) 350 cycles. Investigations into crystalline (Supplementary Fig. 6) and microscopical characteristics show the increase in copper deposition with higher cycles along with smoother and homogenous material deposition (Supplementary Fig. 7) with crystallisation in similar phase as CuNiFe. Further, Supplementary fig. 8 shows the electrochemical activity comparison of different materials electrodeposited for different time (cycles). From the activity comparison, it is obvious 300 cycles of pulses ON and OFF deposited highly active CuNiFe catalyst.

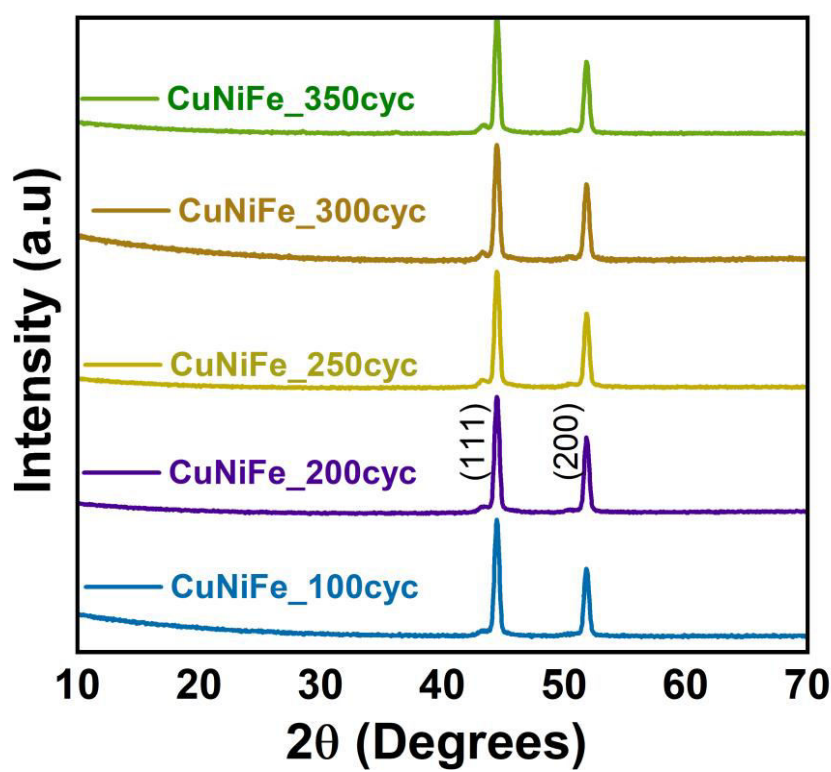

**Supplementary Fig. 6:** XRD graphs of CuNiFe with different mass loading achieved with varying cycles of pulsed deposition (100, 200, 250, 300, 350 cycles). The crystal structure of CuNiFe films deposited does not change, also suggesting similar phase deposition on all deposited films.

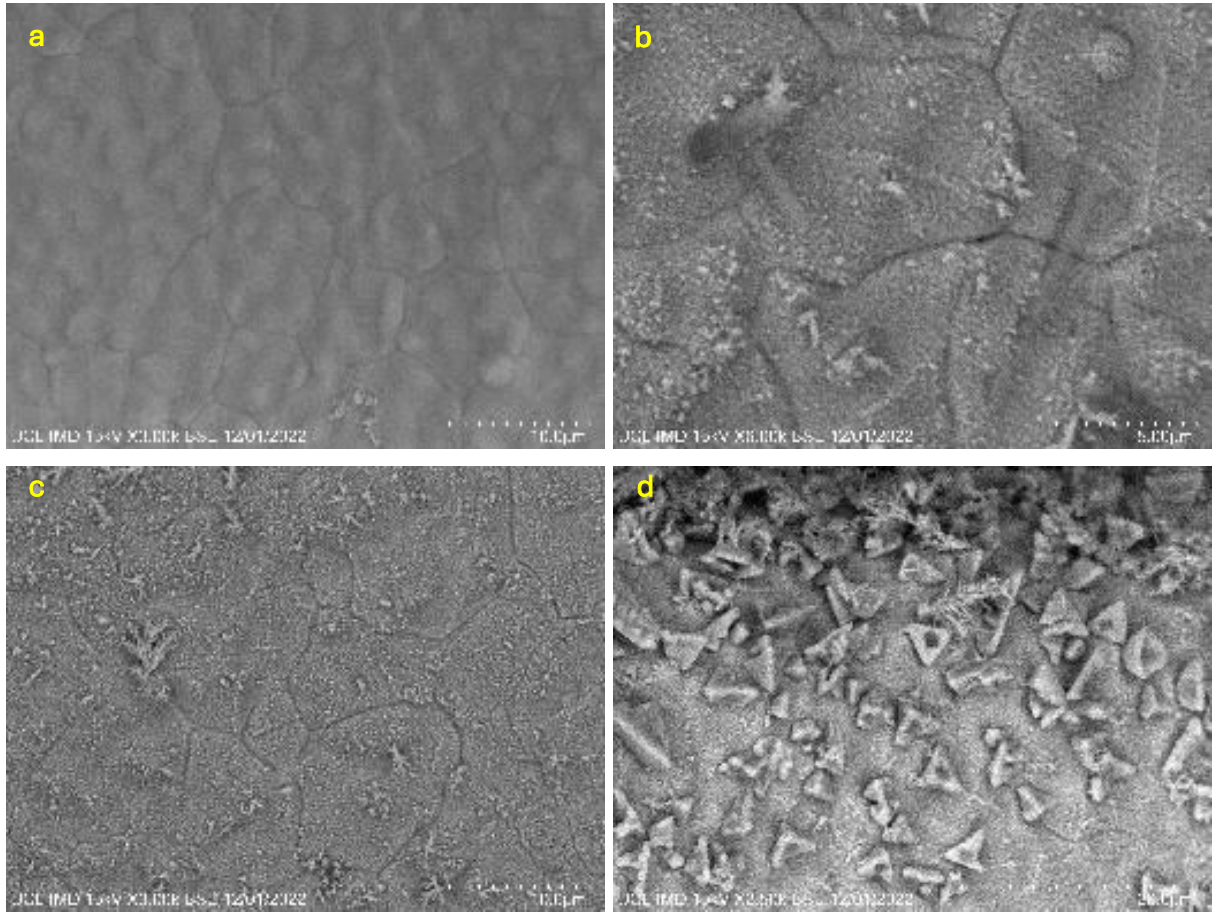

**Supplementary Fig. 7**: SEM images of CuNiFe films with at varying number of total deposition cycles (100-350 cycles). The number of deposition cycles impact the total mass loading in addition to the grain structure. (a) 100, (b) 200, (c) 250, and (d) 350 cycles. As can be seen at 100 cycles od deposition there is very less structuring and of the film with large grains, the grains start appearing made of smaller particles with 200 cycles which further gets refined at 250 cycles. However, at 350 cycles along with smaller particles bigger polygonic structures appear.

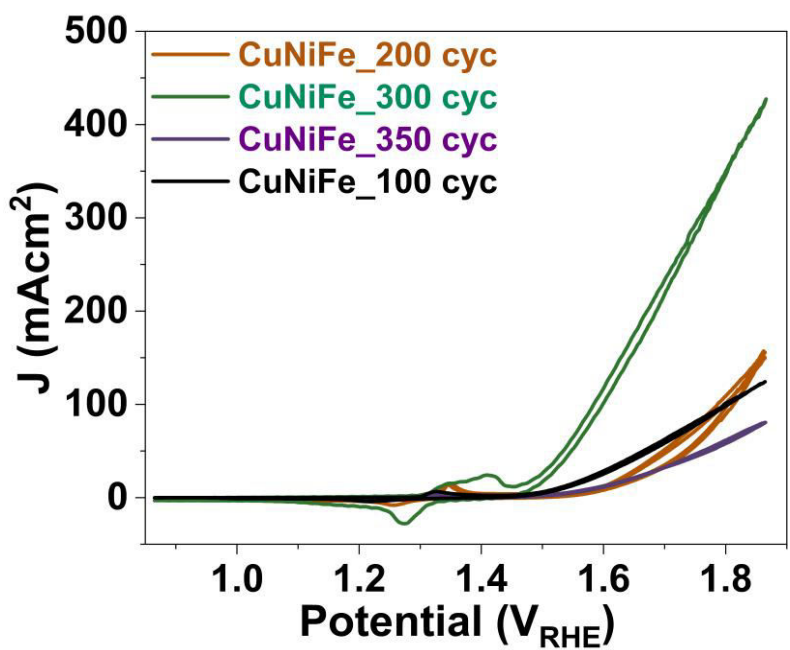

**Supplementary Fig. 8:** CV polarization curves recorded in H-cell containing 1M KOH for materials deposited at varying number of deposition cycles. The optimum mass loading and associated unique microscopical and electronic structure of the sample is recorded at 300 cycles of deposition.

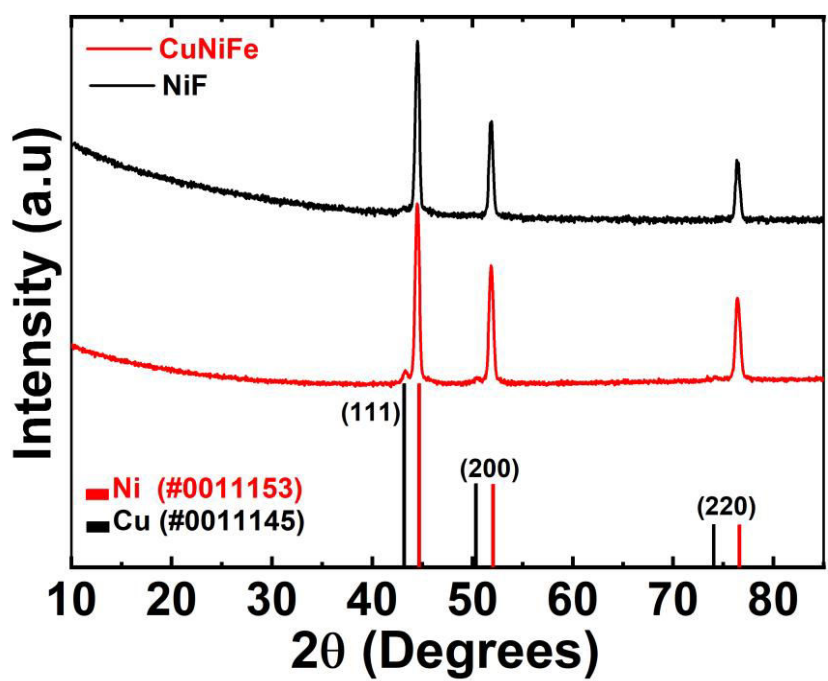

**Supplementary Fig. 9:** XRD traces of Ni foam and CuNiFe film. The peaks corresponding to Cu(0) and Ni(0, II) can be observed at 44.5°, 51.9°, and 76.4° are indexed to (111), (200), and (220) of nickel (PDF #04-0850, whereas additional peaks for Cu (PDF #04-0836) at (111), (200), and (220) planes appear at 43.3°, 50.4°, and 74.1°, respectively.

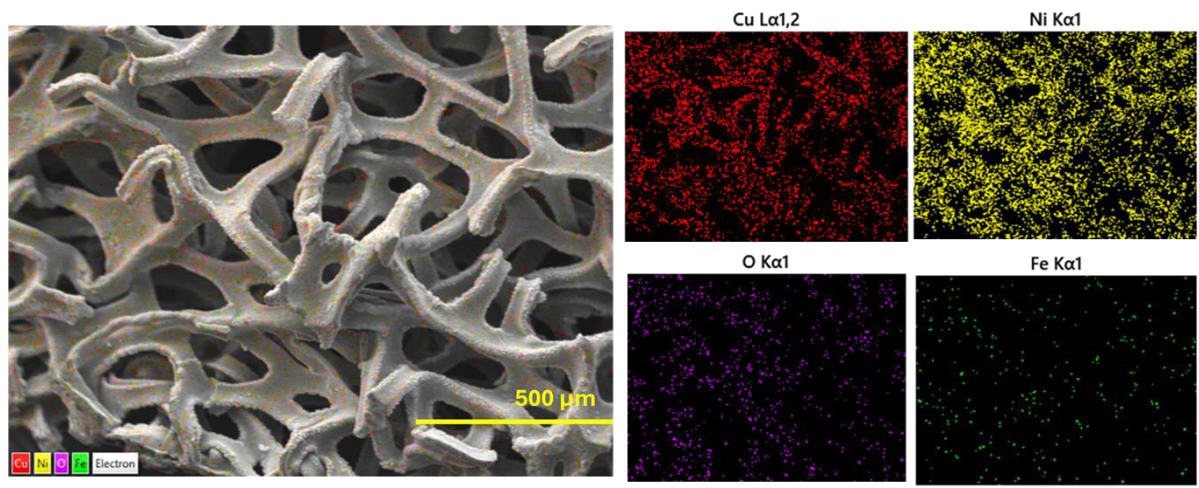

**Supplementary Fig. 10:** Broad range SEM image of CuNiFe and elemental distribution on the surface. As observed all three constituent elements (Cu, Ni, Fe and O (atmospheric and outer oxide shell) are distributed all over the scanned area of ~1.5 mm.

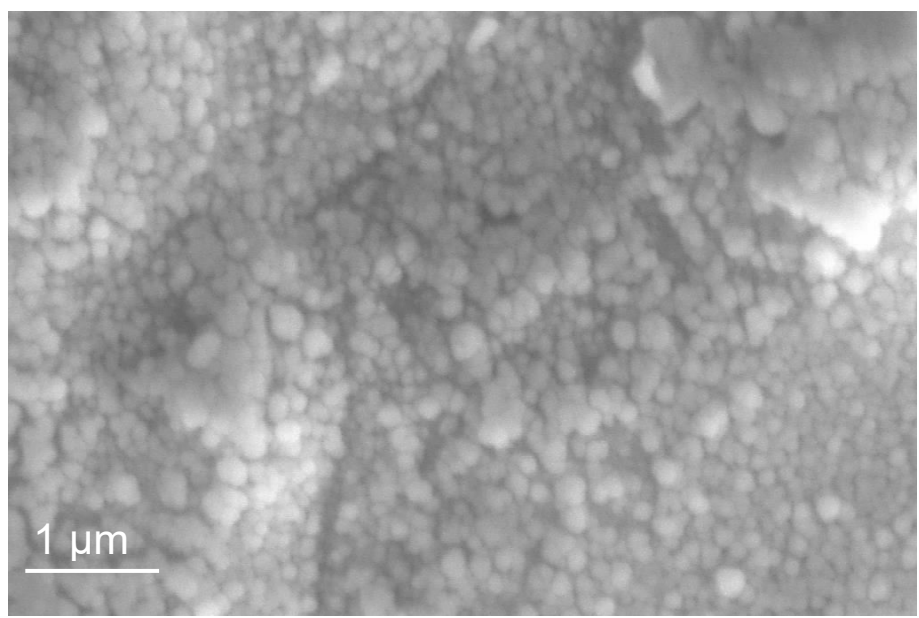

**Supplementary Fig. 11:** High resolution SEM showing particles uniformly distributed on the surface without any pin holes.

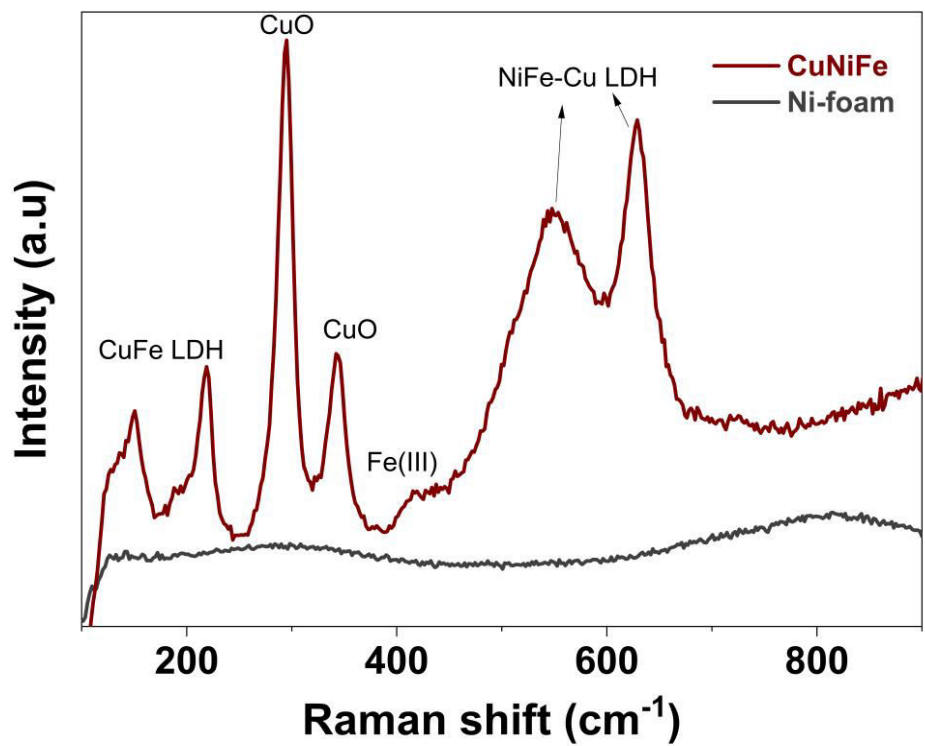

**Supplementary Fig. 12**: Raman spectra of NiF and CuNiFe showing the peaks ascribed to CuFe (100-200cm$^{-1}$), NiFe-Cu LDH (430-700cm$^{-1}$), Fe(III) peak can also be seen around 413cm$^{-1}$, corroborating the unique presence of Fe(III) in the catalyst layer.

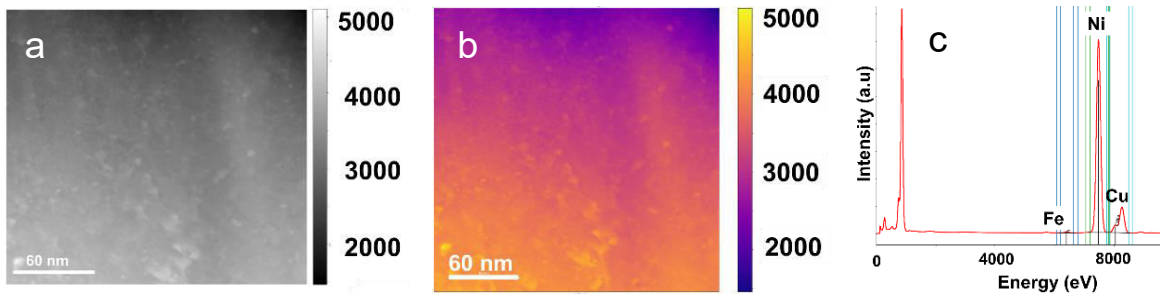

**Supplementary Fig. 13:** HAADF image of the CuNiFe and elemental composition as analysed by FFT. (a) HAADF image of an area on CuNiFe FIB fragment, (b) mixed elemental distribution of Cu, Ni, Fe on the analysed area on the HAADF area, and (c) Energy dispersive spectroscopic (EDS) graph showing energies pertaining to Cu, Ni, and Fe on the analysed area section. The EDS distribution and energies confirm presence of Cu, Ni, Fe showing relative percentage.

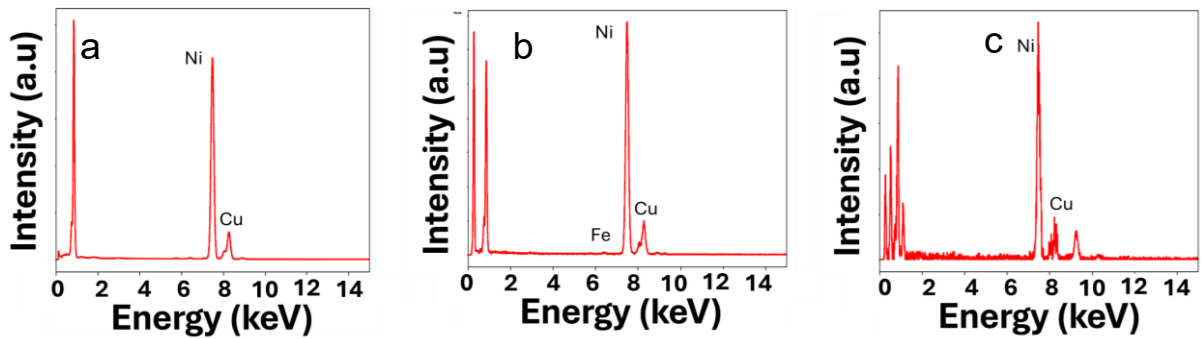

**Supplementary Fig. 14:** Energy dispersive x-ray spectroscopy graphs of Ni (a), Fe (b) and Cu (c) as analysed across cross section of the FIB flagellate.

**Table S1:** Atomic percentage of the constituent elements (cu, Ni, Fe) in CuNiFe and Ni foam samples. The values were generated after the peak fits and have an error of 0.1%.

| Sample name | Cu (%) | Ni (%) | Fe (%) |
|---|---|---|---|
| CuNiFe@Ti | 88.4 | 3.1 | 8.5 |
| CuNiFe@Ni foam | 79.32 | 16.0 | 4.7 |
| Ni foam | 0 | 94.4 | 5.6 |

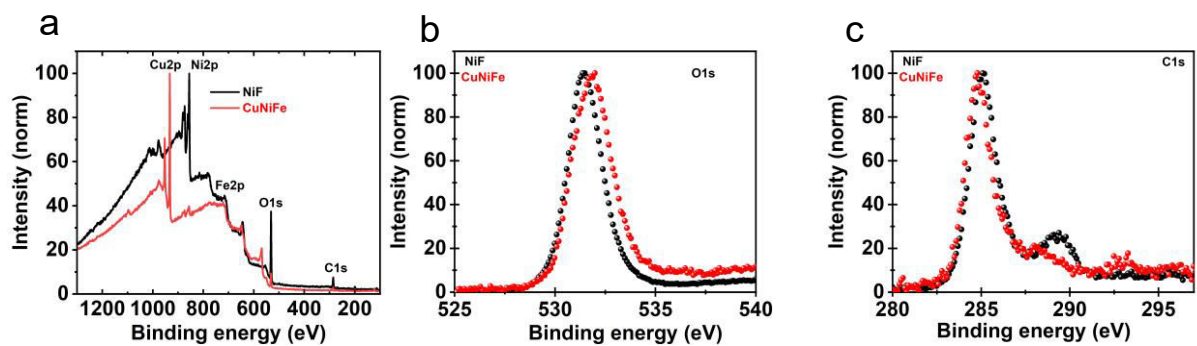

**Supplementary Fig. 15**: XPS spectra of CuNiFe, (a) Survey spectrum showing peaks for Cu, Ni, Fe in CuNiFe and Ni, Fe in Ni-Foam, (b) O 1*s* spectra, and (c) C 1*s* spectra of Ni-Foam and CuNiFe. There is a shift towards higher energies in O 1*s* in CuNiFe owing to presence of more M-O bonds.

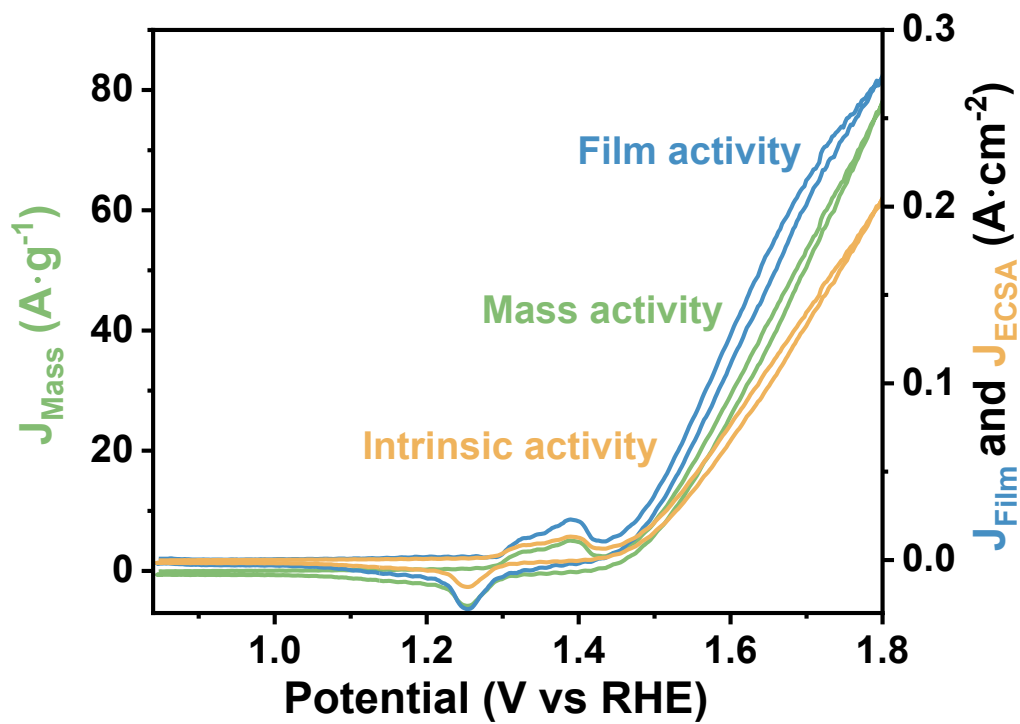

**Supplementary Fig. 16**: CV polarization curves showing mass activity, thin film activity, and specific activity of CuNiFe film on nickel foam substrate.

**Supplementary table S2**: Mass loading of CuNiFe in the optimised catalyst.

| Mass Ni foam before deposition, g | Mass Ni foam after deposition, g | Mass CNF, g |
|---|---|---|
| 0.0884 | 0.0928 | 0.0044 |
| 0.0883 | 0.0928 | 0.0045 |
| 0.0862 | 0.0917 | 0.0055 |
|  | **Avg. mass** | **0.0048** |

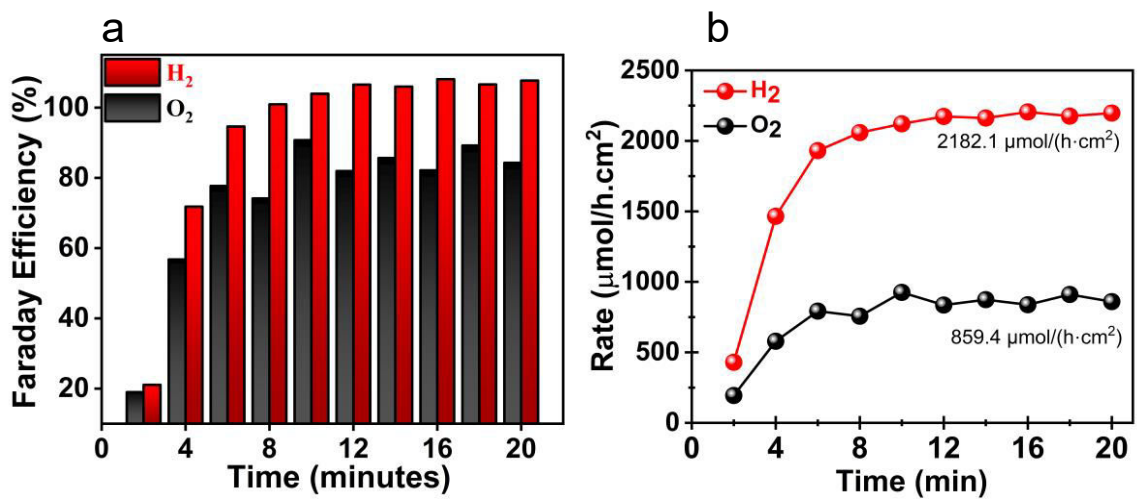

**Supplementary Fig. 17:** Gas evolution quantification of $H_2$ and $O_2$ with time at 100mA/cm$^2$ current in 1M KOH on CuNiFe with in-line gas chromatography, (a) FE, and (b) molar rate production.

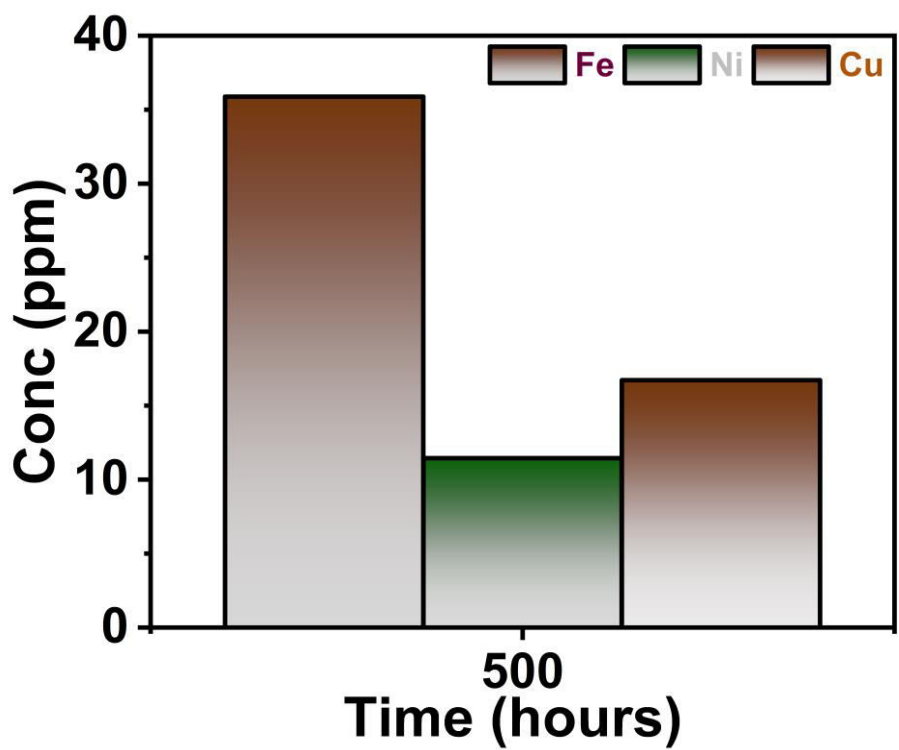

**Supplementary Fig. 18:** ICPMS graph showing the net corrosion or dissolution of Cu, Ni and Fe from the electrode after 500 hours of continous operation.

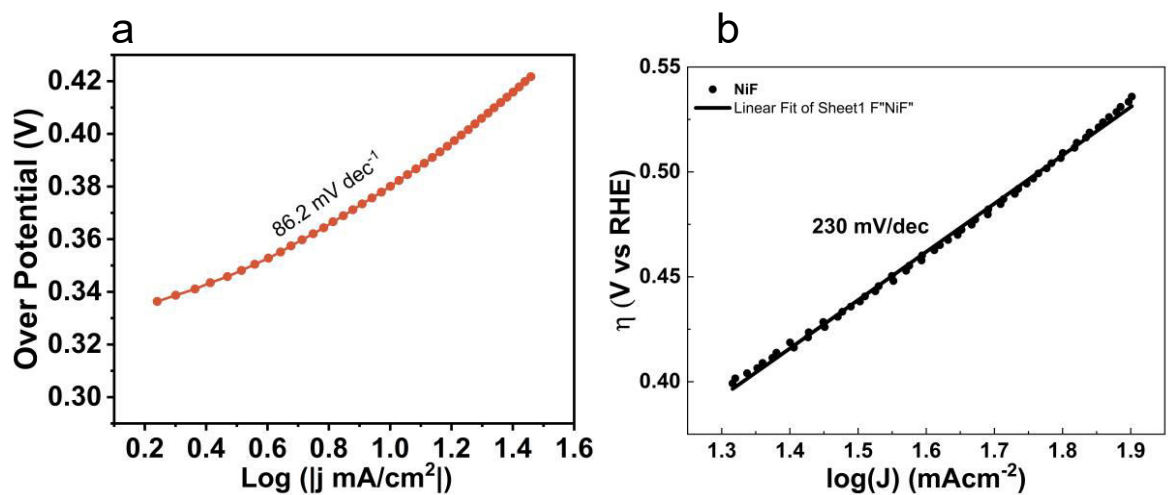

**Supplementary Fig. 19:** Tafel slope of (a) CuNiFe and (b) Ni-foam.

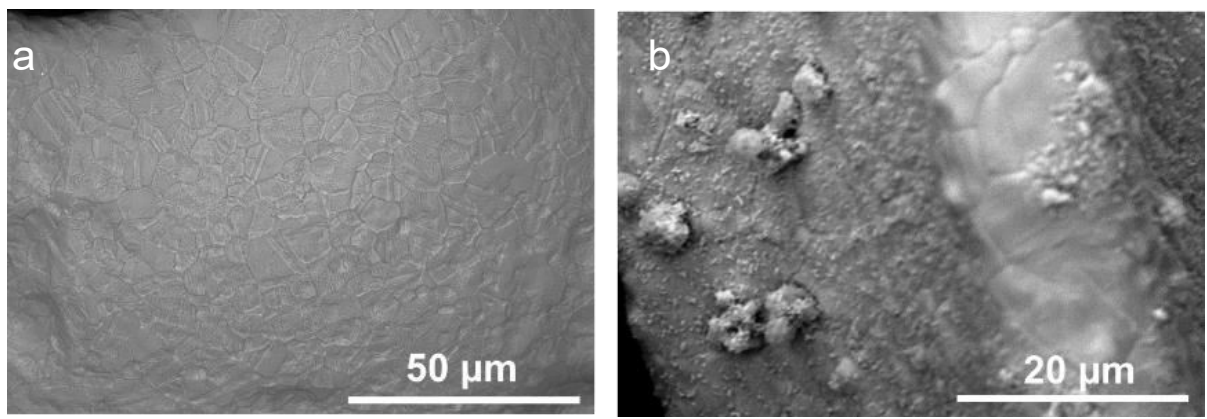

**Supplementary Fig. 20:** Post electrolysis microscopical images of CuNiFe as observed under SEM at 50 μm (a) and (b) 20 μm magnifications. The repetetive grain boundaries can be still observed (a), The particle nature of the film is retained with emergence of resytructured larger clusters (b).

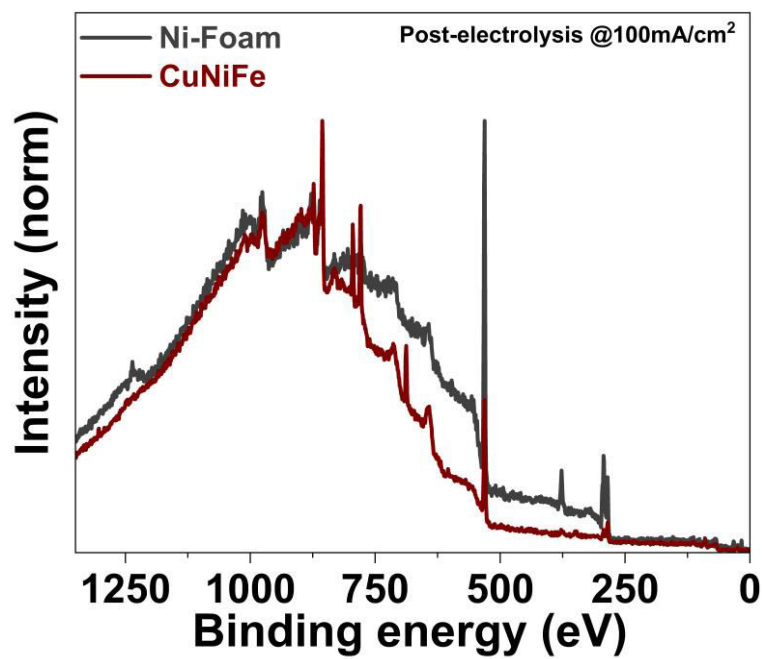

**Supplementary Fig. 21:** Survey spectrum of Ni foam and CuNiFe post electrolysis of 24 hours at 100mA/cm$^2$.

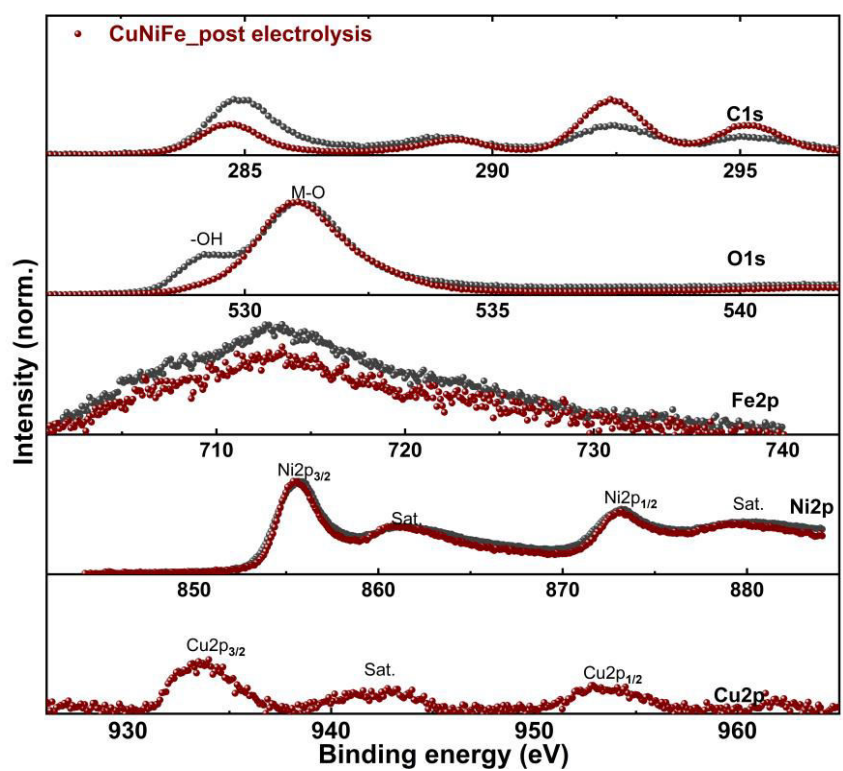

**Supplementary Fig. 22:** Elemental spectrum of Cu, Ni, and Fe in Ni-foam and CuNiFe post electrolysis of 500 hours at 100 mA/cm$^2$. The Cu in CuNiFe is oxidised a expected into (I) and (II) states. The formation of Ni-OOH in Ni-foam is evident in the O1s spectrum while stable M-O bonds are present in CuNiFe O1s after electrolysis.

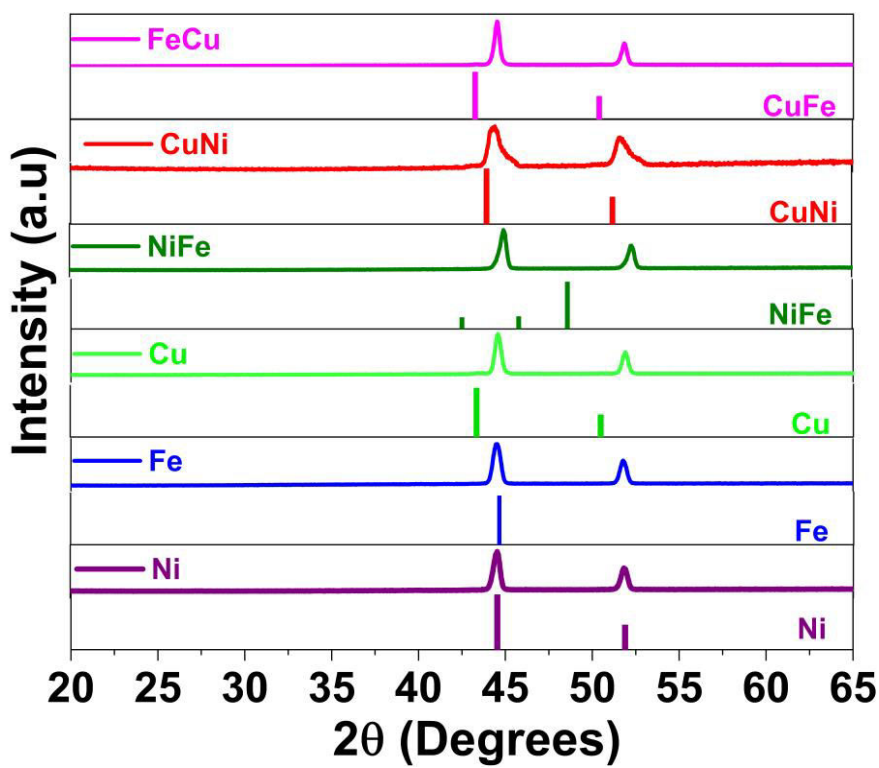

**Supplementary Fig. 23:** XRD graphs of mono-, bi-metallic systems. The bar graph shows the ICCDS patterns for Ni, Fe, Cu, NiFe, CuFe, and CuNi.

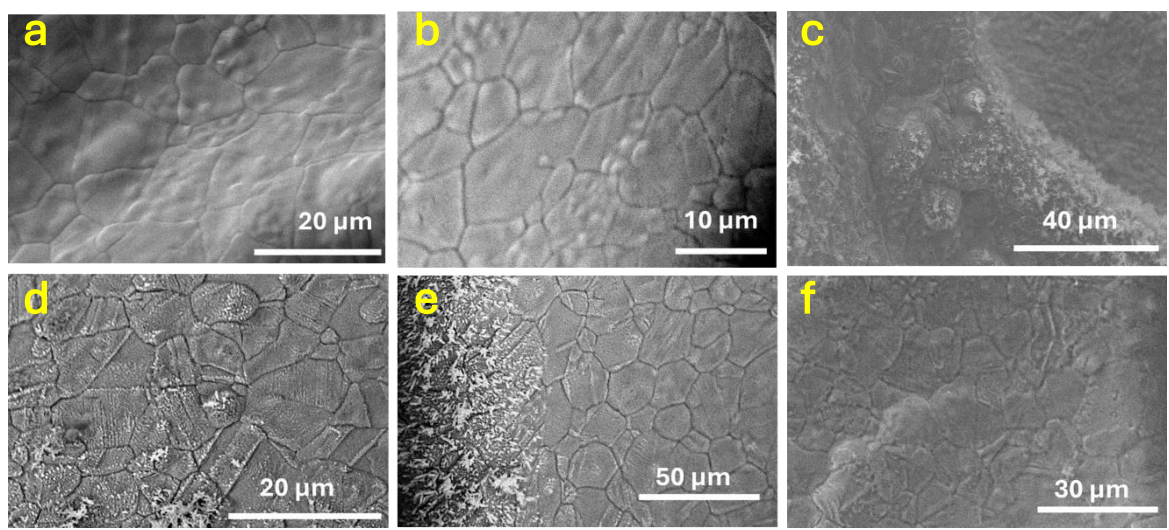

**Supplementary Fig. 24**: SEM images of mono-, and bi-metallic thin films on nickel foam. (a) Ni, (b) Fe, (c) Cu, (d) CuFe, (e) CuNi, and (f) NiFe.

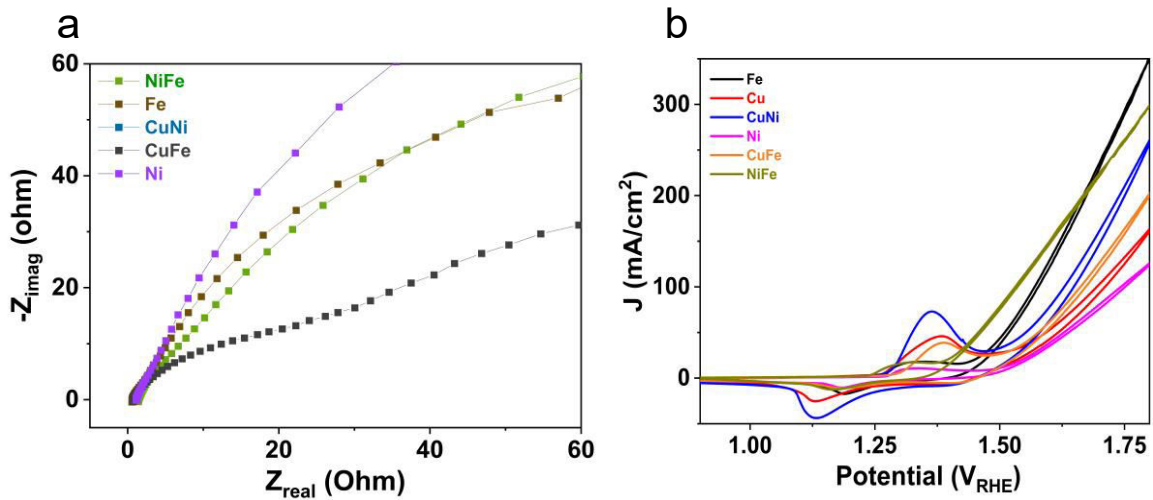

**Supplementary Fig. 25:** (A) EIS at OCP, (b) CV polarization curves at for mono-, and bi-metallic systems systems in 1M KOH in H-cell.

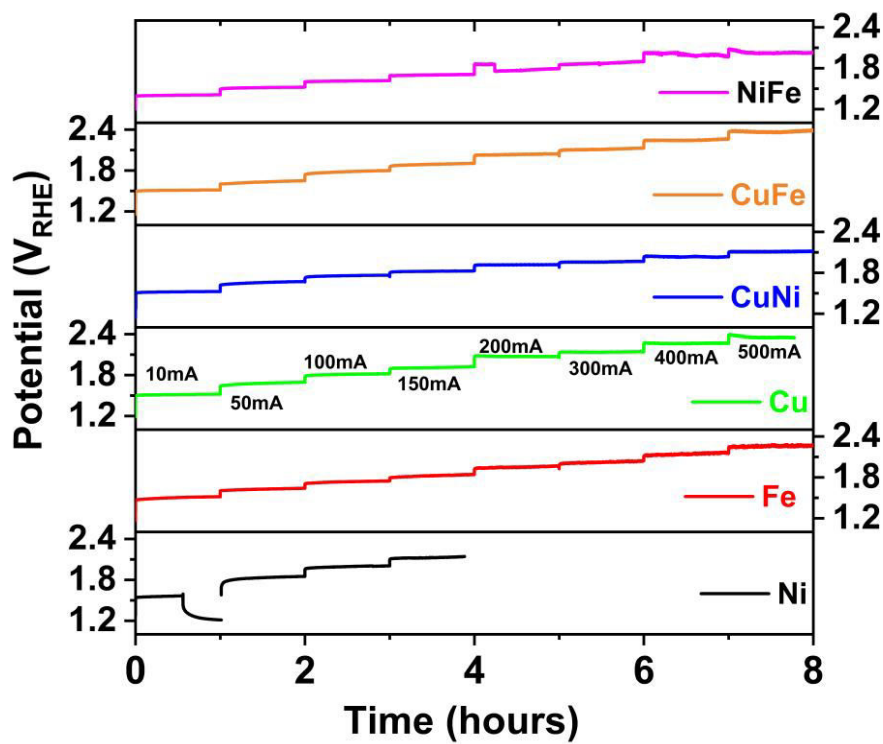

**Supplementary Fig. 26:** Chronopotentiometric polarization curves at different mono- (Cu, Ni, Fe), bi-metallic (CuNi, CuFe, NiFe) systems in 1M KOH at different current dnesities (10-500 mA/cm$^2$). These steady state traces were used to get the overpotentials over transient response of CVs which are usually lower than these.

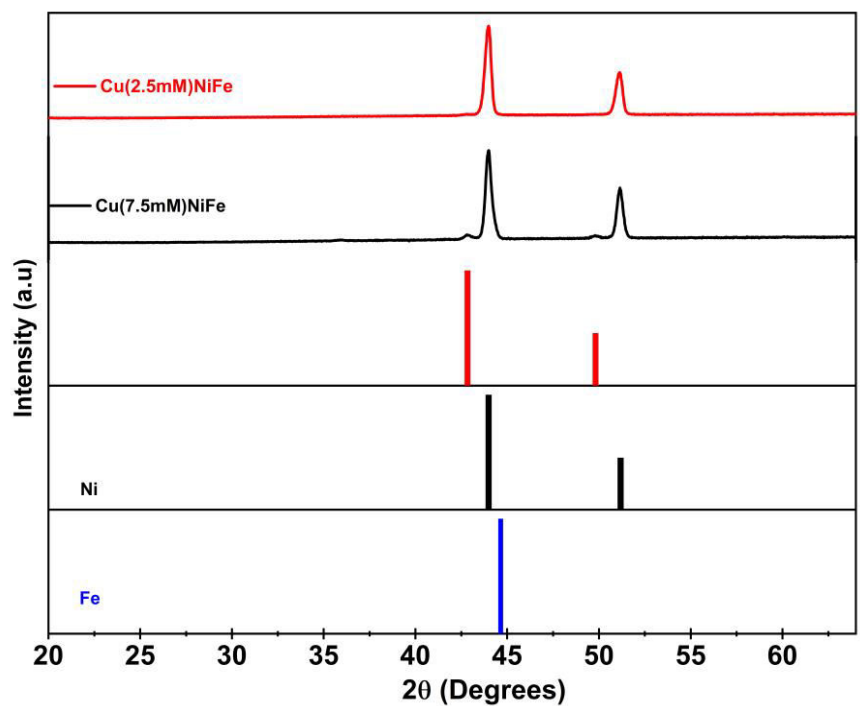

**Supplementary Fig. 27**: XRD images of Cu(2.5mM)NiFe, and Cu(7.5mM)NiFe, showing peaks for Cu, CuNiFe in addition to nickel peaks.

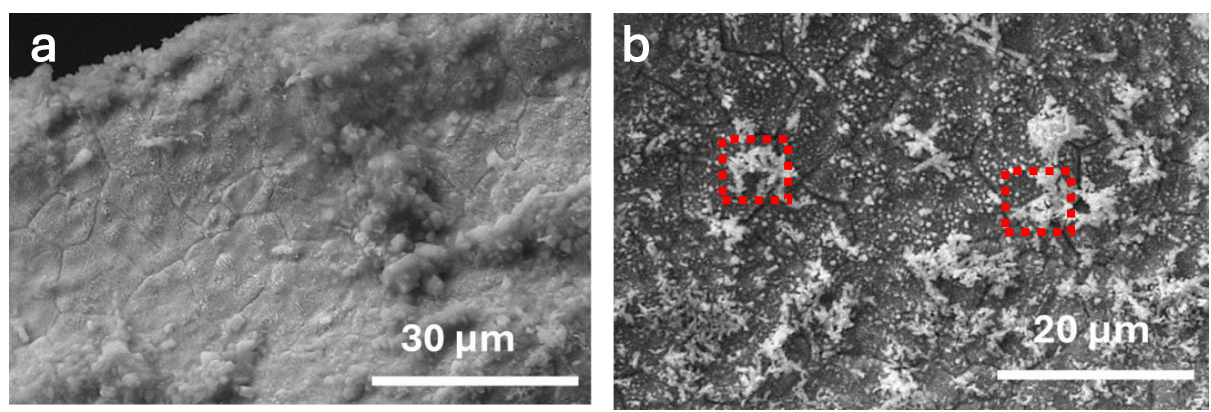

**Supplementary Fig. 28**: SEM images of (a) Cu(2.5mM)NiFe and, (b) Cu(7.5mM)NiFe. In lower concentration of copper there is formation of copper cubes along with adherent layer, due to less dissolution and redeposition competetion from Fe and Ni. While in excess copper concentration a lot of dendrites are formed which increase surface area but undergo huge surface restructuring and dissolution.

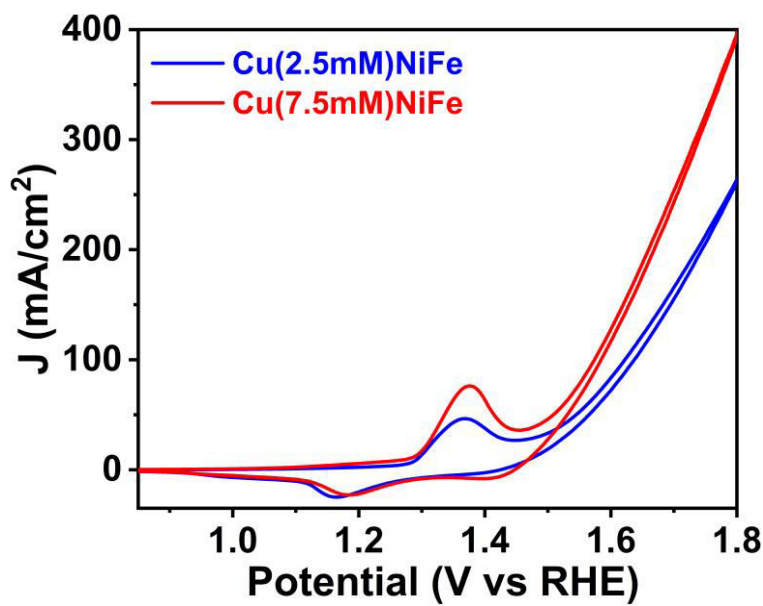

**Supplementary Fig. 29**: CV traces recorded at Cu(2.5mM)NiFe, and Cu(7.5mM)NiFe, showing enhancement in surface area and current. The pre-OER oxidation peak area shows an enhancement suggesting higher number of active sites avaiulable with increasing concentration of Cu in the deposition bath.

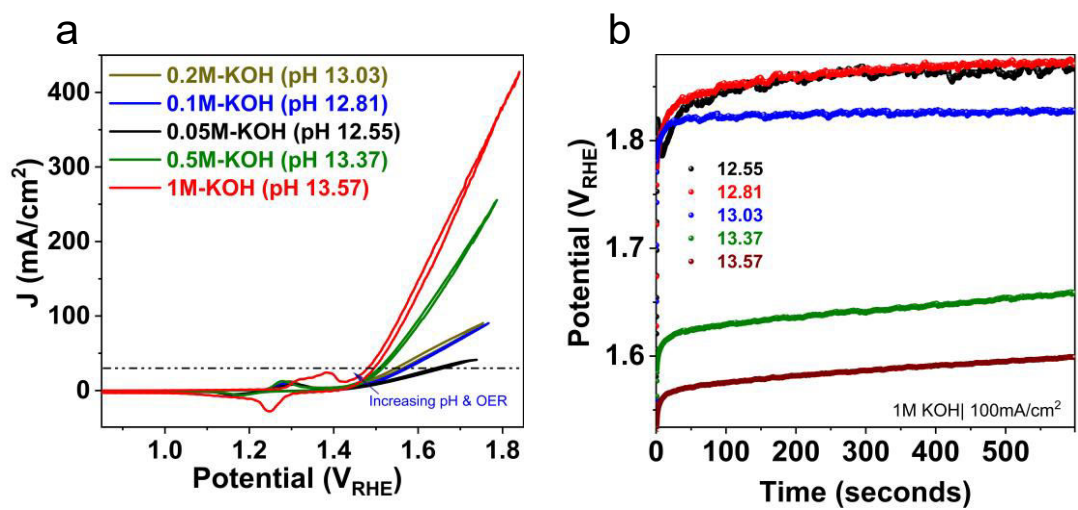

**Supplementary Fig. 30:** (a) CV traces of CuNiFe recorded at different pH (12.55-13.57) for insights into mechnaism, and (b) Chronopotentiometric traces on CuNiFe films tested at different pH by changing Koh concentrations from 0.01 to 1M.

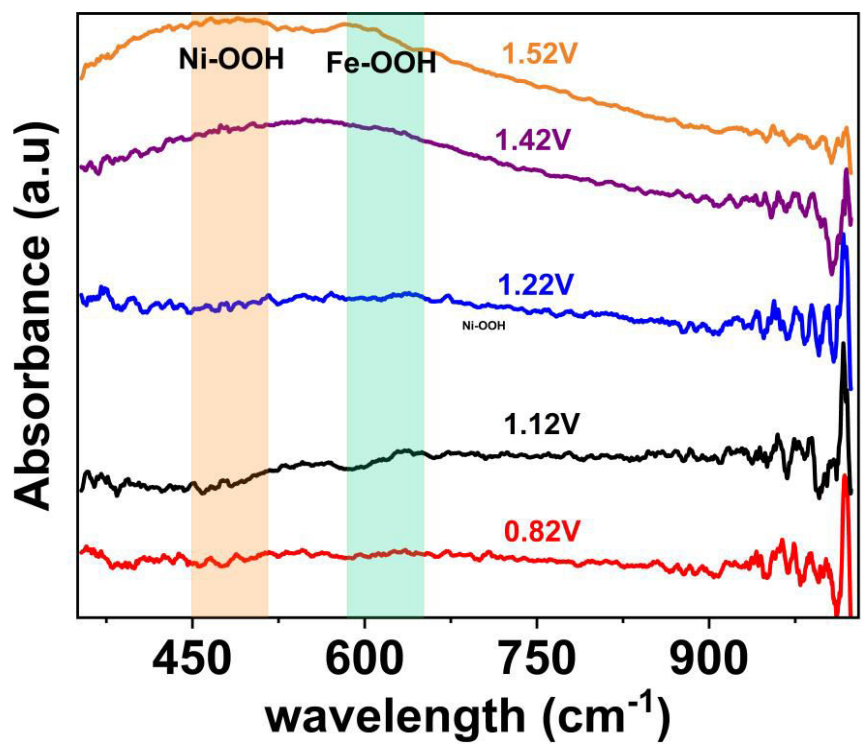

**Supplementary Fig. 31:** In-situ UV-vis spectra of Ni-Foam recorded in reflection mode on Spellec metrohm in 1M KOH at different potentials. There are only two peaks observable after at 1.42V pertaining to Ni-OOH and Fe-OOH formation. There are no peaks or signal aroun 700-900 nm which has been assigned due to formation of Cu-FeNi-OOH species as observed on CuNiFe.[11,12]

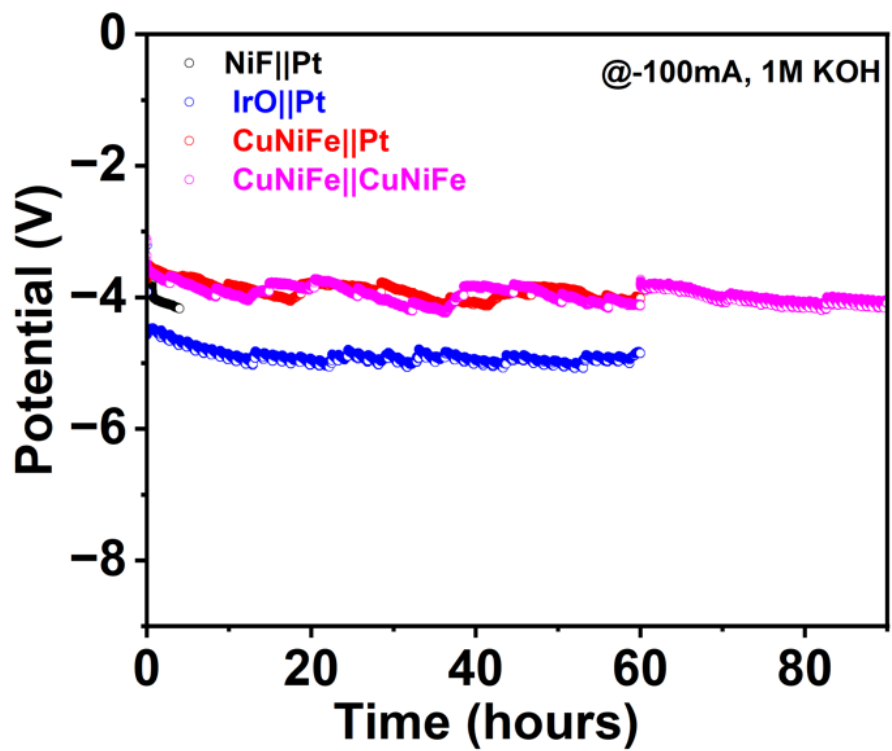

**Supplementary Fig. 31**: Alkaline water electrolysis full cell (H-Cell) at –100 mA/cm$^2$ with Pt as cathode and different anodes (IrO, Ni-foam, and CuNiFe). Further, a symmetrical electrode pair from CuNiFe was made and tested in 1M KOH (H-cell). CuNiFe shows comparable hydrogen evolution activity to Pt.

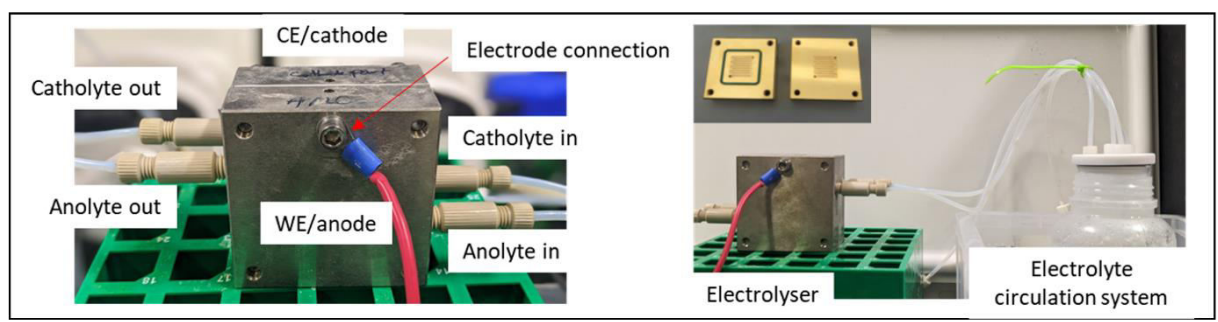

**Supplementary Fig. 32**: Images of the zero-gap electrolyser cell used for AWE and AEM-WE measurements in 30% KOH, 25-60 °C temperature and 1A/cm$^2$ currents. The image also shows the electrolyte circulation.

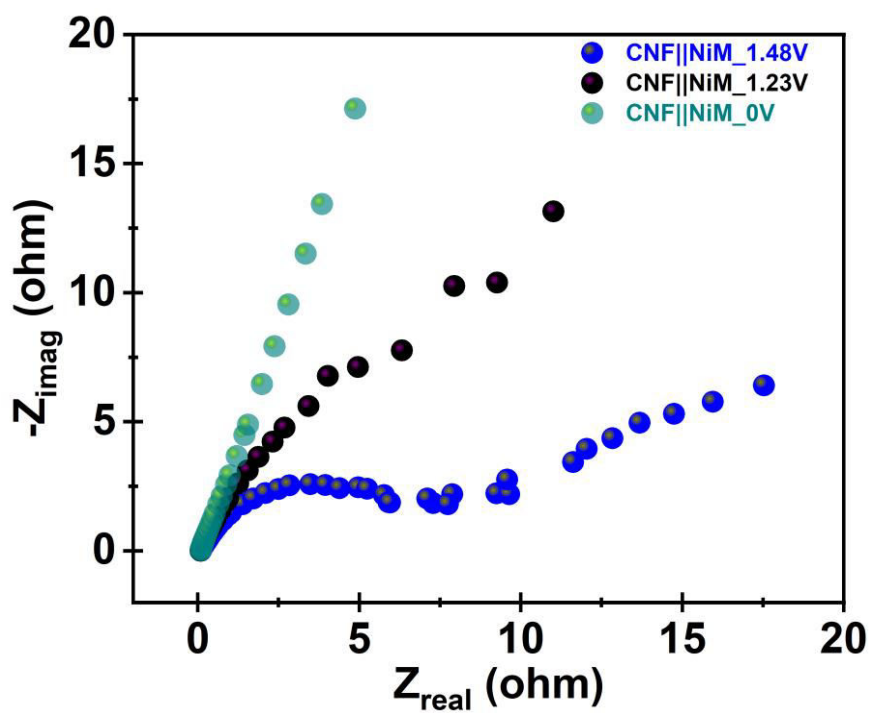

**Supplementary Fig. 33** Potentiostatic EIS of CuNiFe at 0V, 1.23V, and 1.48V.

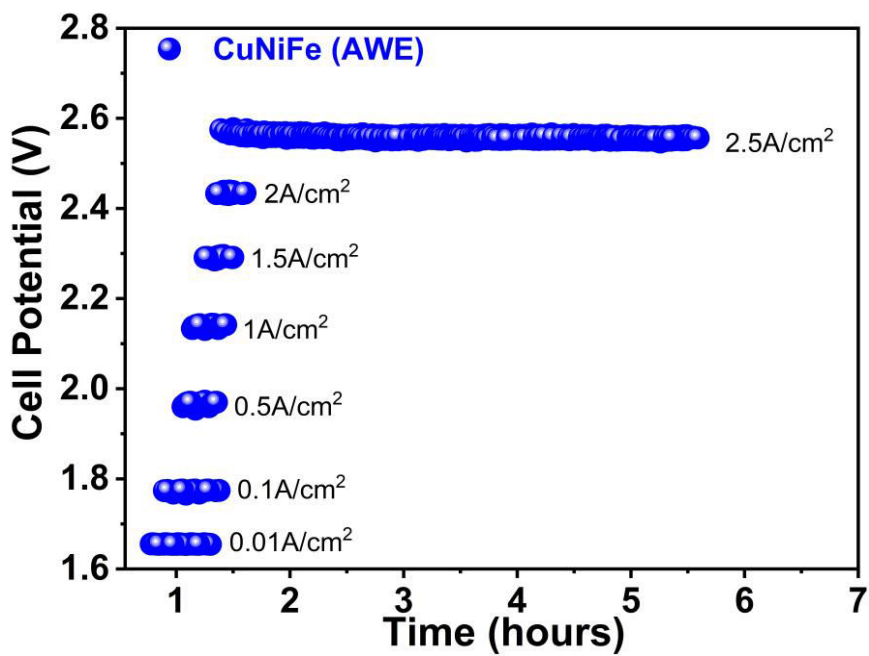

**Supplementary Fig. 34**: Chronopotentiometric traces of CuNiFe in AWE configuration at different applied currents (10- 2500 mA/cm$^2$).

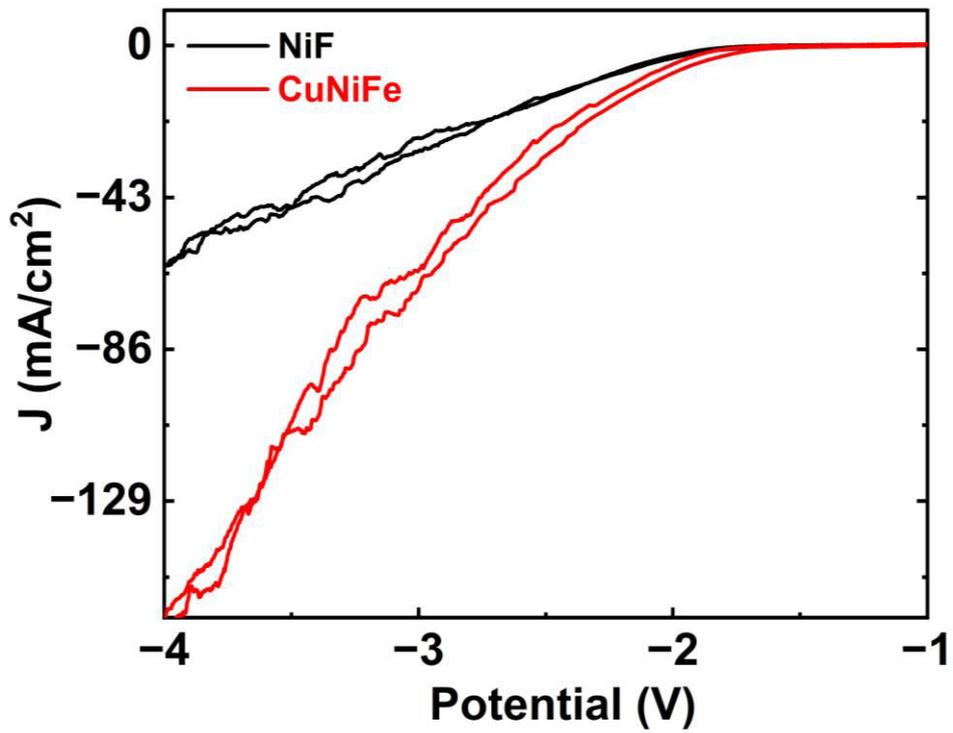

**Supplementary Fig. 35:** CV polarization curves on 40% Cu-vulcan in 1M KOH on CuNiFe and NiF as anodes in zero gap on 25cm$^2$ electrode.

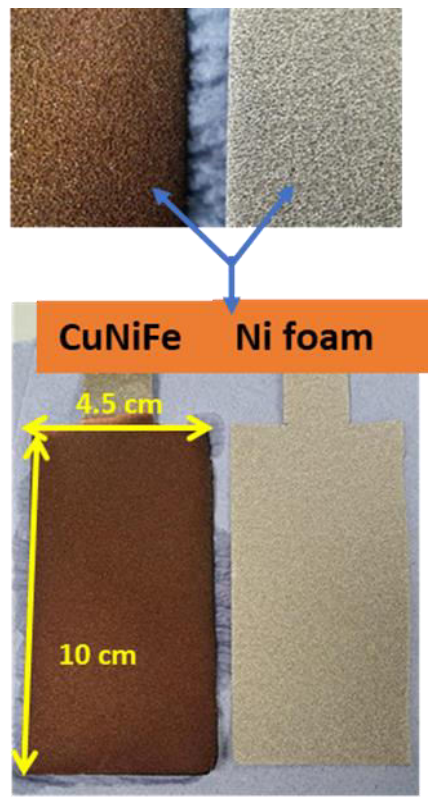

**Supplementary Fig. 36:** macro digital image of deposited CuNiFe (50 cm$^2$) on NiF. Also is shown the random 5 pieces extracted from the electrode.

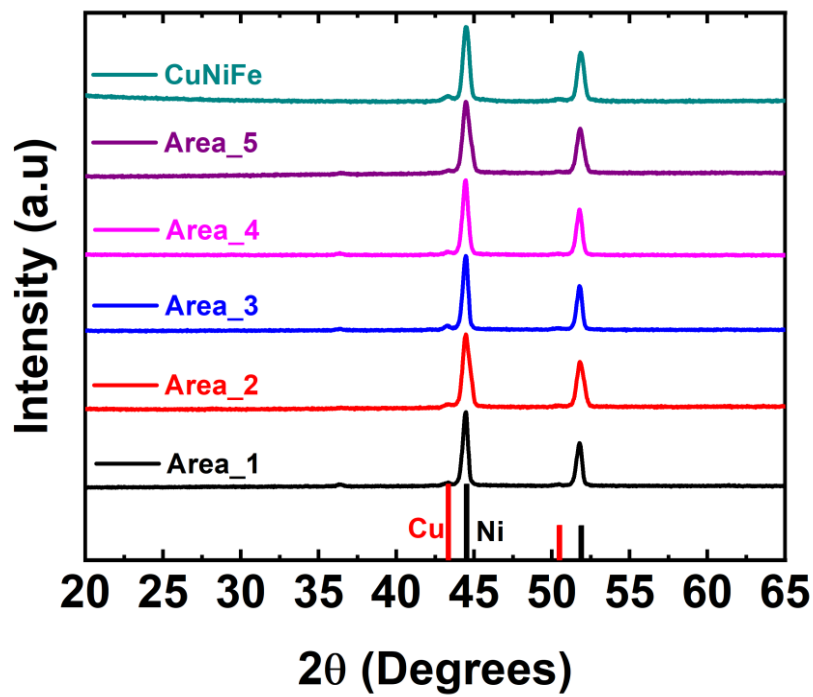

**Supplementary Fig. 37:** XRD graphs observed on the five extracted areas from the large CuNiFe electrode. The patterns are similar to 1-4 cm$^2$ CuNiFe.

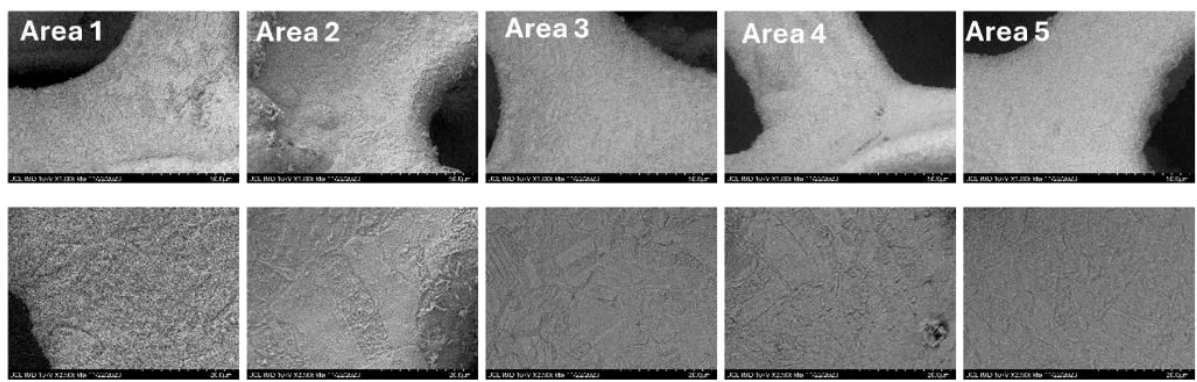

**Supplementary Fig. 38:** SEM images recorded at five different randomly extracted 1 cm$^2$ area from 50 cm$^2$ fabricated electrode. (Area 1, 2, 3, 4, 5) Upper panel shows the microscopical properties at low resolution (50 µm) and below panel shows the repeatability and conformality at higher resolution (10 µm) throughout the surface.

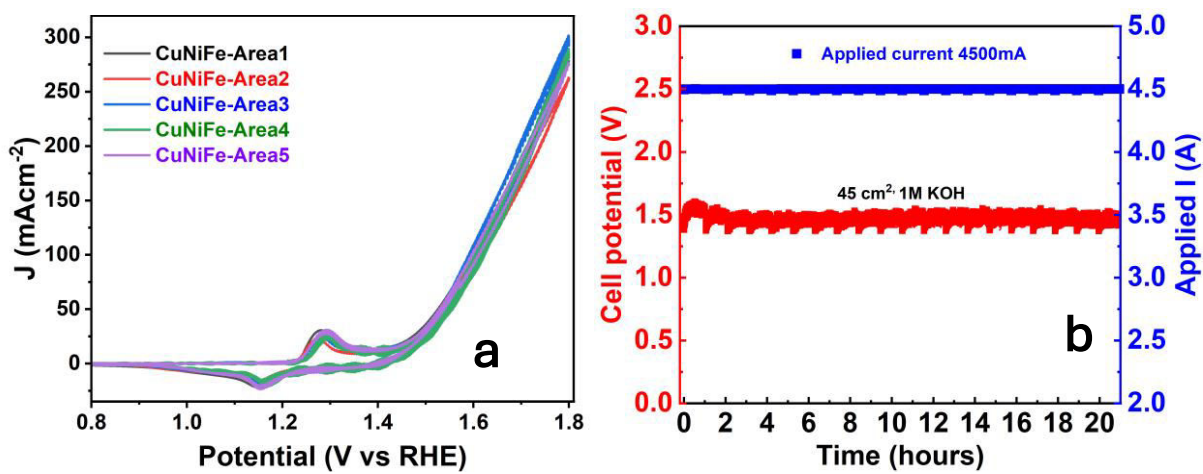

**Supplementary Fig. 39** (a) CV traces on different areas extracted from the large electrode showing similar currents and activity towards OER, and (b) Chronopotentiometric measurement on CuNiFe of 45 cm$^2$ showing similar overpotential to 1 cm$^2$ electrode and stability for 20 hours with 20 ON OFF.

**Supplementary note 4**

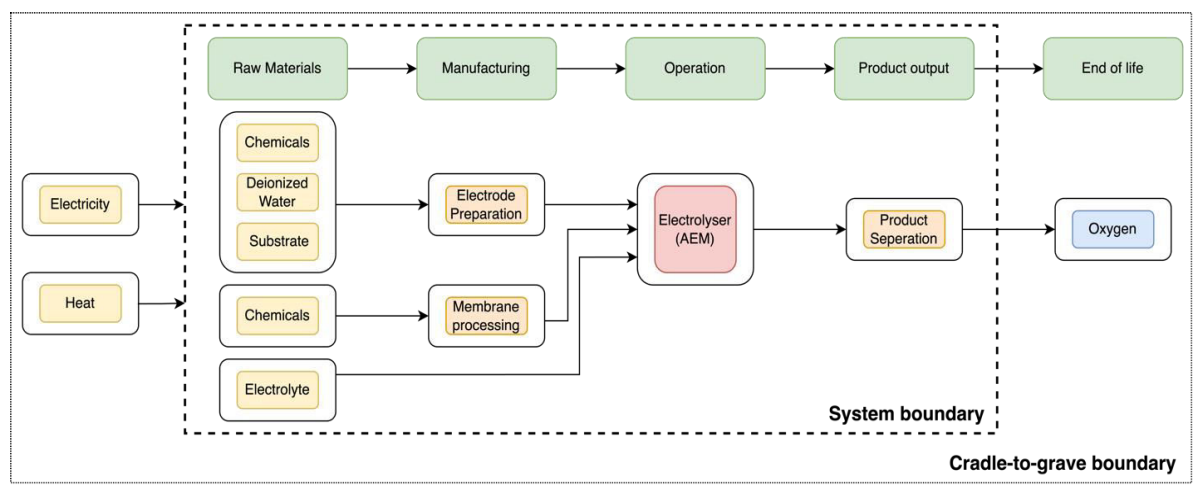

**Supplementary Fig. 40:** Cradle-to-gate system boundary for producing hydrogen and oxygen by electrochemical system AEM cell.

In particular, the AEM cell comprises a platinum mesh cathode, an anionic exchange (AEM) membrane, and the comparative anode catalysts with 1 cm$^2$ reaction area. Since this study focuses on the anodic oxygen evolution reaction, related products such as cathodic catalysts and hydrogen are excluded from the study. During the reaction, electrolyte (1 M KOH) was pumped into the AEM cell, where it was subsequently split into hydrogen and oxygen at the cathode and anode, respectively, driven by an external power supply. In this process, it is assumed that the selectivity of the oxygen product is 100% and the operational energy efficiency is 90%. The electricity is supplied by the United Kingdom power grid, sourced from a mix of energy resources, with the composition calculated based on statistics from the IEA. Data sources for synthesis energy, materials, and operation conditions were obtained from experimental measurements and relevant literature databases as shown in Supplementary table 3. The estimation of electrode and catalyst usage was based on their performance, including lifetime and activity. The environmental impact was assessed using SimaPro with the Ecoinvent v3.8 database on a Europe scale.[13] For environmental impact calculation, carbon footprint is calculated according to IPCC GWP 100a method, and the CML-IA baseline method is applied to evaluate 10 impact categories: Abiotic depletion (elements) (ADPe), Abiotic depletion (fossil fuels) (ADPf), Ozone layer depletion (ODP), Human toxicity (HT), Freshwater aquatic ecotoxicity (FWAEC), Marine aquatic ecotoxicity (MAEC), Terrestrial ecotoxicity (TEC), Photochemical oxidation (PCO), Acidification (AP), Eutrophication (NP). in addition, energy consumption was calculated by Cumulative Energy Demand method. [14]

**Supplementary table 3:** AEM operating conditions for the production of 1 kg $O_2$ from CuNiFe and $IrRuO_2$.

| Category | CuNiFe | $IrRuO_2$ | Unit |
|---|---|---|---|
| Current Density | 100 | 100 | mA/cm2 |
| Cell voltage | 1.5 | 1.7 | V |
| Lifetime | 500 | 100 | hours |
| Operation time | 33501 | 33501 | hours |
| Area Require | 67 | 335 | cm2 |
| Energy consumption | 5025.3 | 5695.3 | Wh |